\documentclass{article}
\usepackage{setspace}
\usepackage{PRIMEarxiv}
\usepackage{natbib}
\usepackage[utf8]{inputenc} 
\usepackage[T1]{fontenc}    
\usepackage{url}            
\usepackage{booktabs}       
\usepackage{amsfonts}       
\usepackage{nicefrac}       
\usepackage{microtype}      
\usepackage{lipsum}
\usepackage{fancyhdr}       
\usepackage{graphicx}       
\usepackage{bm}
\usepackage{multirow}
\usepackage{adjustbox}
\usepackage{float}
\usepackage{parskip}
\usepackage{algorithm,algpseudocode}
\usepackage{amsthm}
\usepackage{soul}
\usepackage{comment}

\newtheorem{theorem}{Theorem}[section]
\newtheorem{lemma}[theorem]{Lemma}
\newtheorem{definition}{Definition}
\newtheorem{proposition}[theorem]{Proposition}%
\newcommand{\bs}{\mathbf{s}}

\newcommand{\E}{\mbox{E}}

\newcommand{\RV}{\mbox{RV}}
\newcommand{\pr}{\text{Pr}}

\newcommand{\sign}{\text{sign}}
\usepackage[normalem]{ulem}

\usepackage{amsfonts}
\usepackage{graphicx,psfrag}
\usepackage{tabularx}
\usepackage{makecell}
\renewcommand{\arraystretch}{0.8}  
\usepackage{xcolor}
\definecolor{neg}{RGB}{244, 109, 67}
\definecolor{pos}{RGB}{102, 194, 165}
\newcommand{\negg}[1]{\textcolor{neg}{#1}}
\newcommand{\pos}[1]{\textcolor{pos}{#1}}
\usepackage{bbm}
\usepackage{mathtools}
\usepackage[shortlabels]{enumitem}
\usepackage{mdframed}
\usepackage{siunitx}
\usepackage[normalem]{ulem}
\usepackage{float}
\usepackage{lineno}
\usepackage{bm}

\title{Modeling high and low extremes with a novel dynamic spatio-temporal model}

\author{
  Myungsoo Yoo\thanks{Author of correspondence: myungsoo.yoo@austin.utexas.edu} \\
  The University of Texas at Austin \\
   \And
  Likun Zhang\\
  University of Missouri \\
   \And
  Christopher K. Wikle\\
  University of Missouri \\
   \And
  Thomas Opitz\\
  INRAE \\
}

\begin{document}
\maketitle

\begin{abstract}
Extreme environmental events such as severe storms, drought, heat waves, flash floods, and abrupt species collapse have become more prevalent in the earth-atmosphere dynamic system in recent years. In order to fully understand the underlying mechanisms and enhance informed decision-making, a flexible model capable of accommodating extremes is necessary. Existing dynamic spatio-temporal statistical models exhibit limitations in capturing extremes when assuming Gaussian error distributions, whereas the current models for spatial extremes mostly assume temporal independence and are focused on joint upper tails at two or more locations. Here, we introduce a new class of dynamic spatio-temporal models that capture both high and low extremes using a mixture of heavy- and light-tailed distributions with varying tail indices. Our framework flexibly identifies extremal dependence and independence in both space and time with uncertainty quantification and supports missing data prediction, as in other dynamic spatio-temporal models. We demonstrate its effectiveness using a large reanalysis dataset of hourly particulate matter in the Central United States.
\end{abstract}


\section{Introduction}
\label{sec:1}
Environmental processes often exhibit extreme event episodes with complex spatio-temporal structure and dynamics, in many cases resulting in catastrophic damage to ecosystems and communities, such as increased mortality from PM$_{2.5}$ exposure \citep{Aaron, Evaluating} and economic losses from tropical cyclones \citep{NormalizedHurricaneDamageintheUnitedStates19002022}. Such extreme events have become increasingly common in recent years \citep{doi:10.1073/pnas.1920849117}, and climate change is expected to lead to more frequent extreme variations in weather \citep{Increasing}. 

Given their increasing frequency and potential for severe impact, various measures and indices have been proposed to identify extreme events and support risk management, including the Climate Extremes Index \citep{IndicesofClimateChangefortheUnitedState, Easterling1999} and the Extreme Forecast Index \citep{https://doi.org/10.1256/qj.02.152}. While widely used, these indices are heuristic rather than based on formal statistical models and do not quantify the uncertainty associated with extreme events. 

In statistics, extreme value theory and its applications provide a foundation for modelling extreme, high-impact events. The classic block-maxima and peaks-over-threshold paradigms \citep{Coles2001} give rise to max-stable processes \citep[e.g.,][]{fabd6366-10fa-3a57-9874-e8b1be7d559e} and generalized Pareto processes \citep[e.g.,][]{ferreira2014}, whose appeal lies in their rigorous asymptotic guarantees. To better capture realistic sub-asymptotic extremal dependence, particularly in spatial context, researchers have developed random scale mixture models \citep[e.g.,][]{OPITZ20161,HUSER2017166, huser2019modeling} and max-id models \citep[e.g.,][]{10.1214/12-AOAS591,Opitz2013,Bopp02012021}; see \citet{huser2022advances} for a recent review. Despite these advances, most existing frameworks assume spatial and temporal stationarity; that is, the extremal dependence is identical at fixed space–time lags and does not evolve over time. Such an assumption can be overly restrictive for large domains or under changing environmental conditions, where extremal dependence exhibits pronounced non‑stationarity and typically varies with spatial scale, ranging from near‐independence at long distances to strong dependence at short distances.

In studying extreme events in spatio-temporal contexts, it is important to account for both spatial and temporal dependence, and various models have been proposed to this end \citep[e.g.,][]{10.1214/14-AOAS766, d6a73349-fcc4-3958-ae5c-c4d87fdacd7b, bacro2020, delloro, zhang2024leveraging}, wherein temporal dependence is typically handled by including temporal covariates or allowing model parameters to vary over time, without explicitly modelling the underlying dynamics. Justified by the fact that block‑maxima and peaks‑over‑threshold frameworks concentrate on tail events rather than the full continuous‑time trajectory, the latent spatial process is usually taken to be temporally independent, an assumption that can overlook meaningful serial dependence and dynamic behaviour among extreme occurrences across fine time scales.

To address these gaps, we propose a spatio-temporal dynamic model \citep[e.g., see the reviews in][]{cressie,wikle2019} that captures spatial and temporal dependence while incorporating a regime-switching process for the innovations (or forcing). This framework is physically motivated, as it reflects the influence of external drivers that can result in extreme events \citep[e.g., see][for a discussion of the temperature and precipitation extremes caused by sea surface temperature and sea ice concentration forcing]{DITTUS20181}. Moreover, our model enables the identification of extreme events across space and time, particularly the timing and location of their onset, while also quantifying the associated uncertainty of these events.

Although our focus is on environmental processes, extreme‑event detection is important in other domains.  In finance, for instance, \citet{https://doi.org/10.1002/for.3019} model sudden market moves or ``jumps'' as the product of two random variables. Our approach differs by modelling switches between regimes rather than discrete multipliers. Similarly, stochastic volatility models in econometrics have also been used to characterize extremal behaviour \citep[e.g.,][]{jones2003dynamics,davis2009extremes}. However, such models do not explicitly define extreme behaviour with uncertainty and have not been applied to high-resolution spatial data. By contrast, our framework delivers a unified, uncertainty‑aware treatment of extremes in the rich spatio-temporal databases typical of environmental applications.

The primary contribution of our work is as follows. First, our model overcomes key limitations of recent spatial extreme models. It introduces the first framework that jointly captures flexible spatial, temporal, and space–time interactive extremal dependence  within a single spatio-temporal dynamic specification. We also examine its properties, most notably its extremal dependence structure, supported by theoretical results. Second, our model introduces a model-based approach to detect and identify extreme events in space and time, along with their associated uncertainty. Third, our model can still reliably detect extreme events even under model misspecification with associated uncertainty. We demonstrate the utility of our model using hourly $\text{PM}_{2.5}$ measurements from the Central United States from 19 to 30 March 2024.

This paper is organized as follows. Section~\ref{sec:2} reviews distributions with tails heavier than Gaussian ones, focusing on the variance-gamma and stable distributions as key components of our model. Section~\ref{sec:3} introduces the dynamic spatio-temporal model and the regime-switching process for extreme innovations. Theoretical properties are presented in Section~\ref{sec:theo_res}, followed by the Bayesian hierarchical formulation in Section \ref{sec:Bayes_inference}. Sections~\ref{sec:5} and \ref{sec:6} present numerical experiments and a data analysis. Section~\ref{sec:7} concludes with a discussion.


\section{Background on non-Gaussian innovation distributions}
\label{sec:2}
The ubiquitous Gaussian distribution has light and symmetric tails, with probability mass concentrated around the mean. Its rapid tail decay makes it unsuitable for data with extreme values when these can fall relatively far from the central part of the distribution. Similarly, Gaussian processes have light joint tails and fail to capture tail dependence \citep{davison2012}. Using heavier-tailed distributions can mitigate these drawbacks. Two families are especially useful: first, regularly varying distributions (e.g., the stable family), which possess power‑law tails capable of accommodating very large values; and second, exponential‑tailed yet over‑dispersed distributions (e.g., the variance‑gamma family), whose tails remain heavier than Gaussian while retaining finite moments. Considering linear combinations of heavier-than-Gaussian distributions leads to stronger extremal dependence, including asymptotic tail dependence \citep{Engelke2019}. We provide a brief overview here and direct readers to \citet{nolan2020univariate} and \citet{fischer2023} for comprehensive treatments for stable and variance-gamma distributions, respectively.

 \begin{definition}[Regularly varying distributions ($\mathrm{RV}^{\infty}_{\lambda}$ and $\mathrm{RV}^{-\infty}_{\lambda}$)]
A probability distribution $F$ with upper endpoint $x^\star = \infty$ is regularly varying with index $\lambda \geq 0$ at infinity if $\overline{F}(tx)/\overline{F}(t) \to x^{-\lambda}$ as $t \to \infty$ for any $x>0$, where $\overline{F}(x)=1-F(x)$. Similarly, if $F$ has a lower endpoint $x_\star = -\infty$ and $F(tx)/F(t) \to x^{\lambda}$ as $t \to -\infty$ for any $x>0$,  $F$ is regularly varying with index $\lambda \geq 0$ at negative infinity. We write $F \in \mathrm{RV}^{\infty}_{\lambda}$ or $F \in \mathrm{RV}^{-\infty}_{\lambda}$ respectively. If $\lambda=0$, then $F$ is said to be \emph{slowly varying}.
\end{definition}

\begin{definition}[Exponentially-tailed distributions ($\mathrm{ET}^\infty_{\lambda}$ and $\mathrm{ET}^{-\infty}_{\lambda}$)]
    A random variable $X$ belongs to the class $\mathrm{ET}^\infty_{\lambda}$, with rate $\lambda>0$, if its tail behaviour satisfies $\overline{F}(t+x)/\overline{F}(t)\rightarrow 
    \exp(-\lambda x)$ as $t\to\infty$ for any $x>0$.
    Similarly, we say $X\in\mathrm{ET}^{-\infty}_{\lambda}$, with $\lambda>0$, if $-X$ is in $\mathrm{ET}^\infty_{\lambda}$, i.e., $F(-x-t)/F(-t)\rightarrow \exp(-\lambda x)$ as $t\to\infty$ for any $x>0$.
\end{definition}

By definition, a random variable $X$ having regularly varying distribution in the upper tail is equivalent to $\log(1+X_+)$ (where $X_+=\max(X,0)$, and where we add $1$ inside the logarithm to avoid singularities when $X_+=0$) having exponential-tailed distribution. Specifically, $F\in\mathrm{ET}^\infty_{\lambda}$ with $\lambda>0$ if and only if $\overline{F}(\log(1+\cdot)) \in \mathrm{RV}^{\infty}_{\lambda}$. In analogy, the connection between $\mathrm{ET}^{-\infty}_{\lambda}$ and $\mathrm{RV}^{-\infty}_{\lambda}$ can also be established via a log-transformation $\log(1-X_-)$. If $X$ has standard Gaussian distribution, then $X^2$ is exponential-tailed ($\mathrm{ET}^\infty_{1/2}$), and $\exp(X^2)$ is in $\mathrm{RV}^{\infty}_{1/2}$.

\subsection{Background on the stable distribution}\label{sec:2.2} 

Stable distributions are a family of probability distributions that satisfy the stable laws and are closed under linear combinations. A random variable $X$ is \textit{stable} if and only if there exists $c_n>0$ and $d_n \in \mathbb{R}$ such that $X_1 + \cdots +X_n \stackrel{d}{=} c_n X + d_n$ where $X_1,...,X_n$ are independent and identically distributed (i.i.d.) samples of $X$. 

The stable family can be parameterized by four parameters, and we write $X \sim \mathcal{S}(\lambda,\kappa,\nu,\delta)$, in which the concentration parameter $\lambda\in (0,2]$ controls tail heaviness, and the skewness parameter $\kappa\in[-1,1]$ determines asymmetry. As $\lambda \rightarrow 0$, the tails become heavier, with skewness either to the left or right depending on $\kappa$. When $\lambda<1$, the distribution can be fully skewed left ($\kappa=-1$) or right ($\kappa=1$) concentrating all mass to one side of $[\delta+\nu \tan (\pi \lambda/2)]$, where $\delta \in \mathbb{R}$ is the location and $\nu >0$ is a scale parameter.  Note that special cases of stable distributions include some familiar distributions such as Gaussian ($\lambda=2$), Cauchy ($\lambda=1,\;\kappa=0$), and L\'evy distribution ($\lambda=1/2$, $\kappa=1$). When $0<\lambda<2$ and $-1<\kappa<1$, stable distributions are regularly varying at both $\infty$ and $-\infty$ with index $\lambda$. Specifically, if $w\sim \mathcal{S}(\lambda, \kappa, \nu, \delta)$, then $\Pr(w>x)\sim \nu^{\lambda} c (1+\kappa)x^{-{\lambda}}$ and $\Pr(w<-x)\sim \nu^{\lambda} c (1-\kappa)x^{-{\lambda}}$, with $c=\sin(\pi {\lambda}/2)\Gamma({\lambda})/\pi$. 
The Gaussian case with $\lambda=2$ is not regularly varying because its tails decay much faster.

\subsection{Background on the variance-gamma distribution}\label{sec:2.1} 

The variance-gamma (VG) distribution also has four parameters and appears in various parameterizations in the literature \citep{madan1990,seneta2004,finlay2008,klar2015,fischer2023}. Here, we use the following density function \citep{madan1990,seneta2004}: 
\begin{align*}
f_{\text{VG}}(x; \xi,\theta,\sigma,\mu) = c(\xi,\theta,\sigma,\mu)\times e^{\frac{\theta(x-\mu)}{\sigma^2}}|x-\mu|^{\frac{1}{\xi}-\frac{1}{2}} K_{\frac{1}{\xi}-\frac{1}{2}}\left( \frac{|x-\mu|\sqrt{2\sigma^2 \xi^{-1} +\theta^2}}{\sigma^2}\right),
\end{align*}
where $K_{\alpha}(\cdot)$ is the modified Bessel function of the second kind of order $\alpha\in\mathbb{R}$, and the distribution includes a location parameter $\mu\in\mathbb{R}$, an asymmetry parameter $\theta\in\mathbb{R}$, a spread parameter $\sigma$, and a shape parameter $\xi$. The upper and lower tails of the VG distribution belong to $\mathrm{ET}^\infty_{\lambda^+}$ and $\mathrm{ET}^{-\infty}_{\lambda^-}$ respectively, where $\lambda^{\pm} = \frac{\sqrt{\theta^2 + \sigma^2} \mp \theta}{\sigma^2}$.

The VG distribution is further known to have constructive representation as a normal mean-variance mixture where the standard Gaussian distribution is a boundary case, such that the VG naturally extends the Gaussian. For more background, see \citet{madan1990,kotz2001,seneta2004}, where \citet{kotz2001} uses the notion of Generalized Asymmetric Laplace distribution.  
\section{Extreme Dynamic Spatio-temporal Model}\label{sec:3}
Dynamic spatio‑temporal models (DSTMs) are indispensable whenever a phenomenon evolves over time, is observed imperfectly in space, and is driven by underlying mechanistic processes (e.g., physics or ecology). They have been widely used in applications ranging from atmospheric sciences, hydrology, oceanography to ecosystem management; see \citet{cressie} for an overview. 

The remainder of this section reviews the classical DSTM formulation and then introduces our extension that accommodates extreme events. Before turning to the classical DSTM framework, we clarify our notation. Boldface letters denote matrices (e.g., $\bm A$) or random vectors. A centred dot “$\,\cdot\,$” is used as a placeholder for the suppressed column or row index so that $\bm A_{i,\cdot}$ and $\bm A_{\cdot,j}$ denotes the $i$-th row and the $j$-th column of $\bm A$, respectively. For a time–varying matrix, we write $\bm A_t$ with $t$ being a discrete time index.  We write $\bm (\bm A_t)_{i,\cdot}$ and $(\bm A_t)_{\cdot,j}$ for, respectively, the $i$-th row and the $j$-th column of $\bm A_t$. Individual entries are written $A_{ij}$ or $A_{ij}(t)$.

\subsection{Dynamic Spatio-Temporal Model}
\label{sec:dstm}
The general dynamic spatio-temporal model (DSTM) can be written as
\begin{equation}\label{eqn:data_model}
    Y_t(\bm s)=\mathcal{M}\{Y_{t-1}(\mathcal{D}); \boldsymbol{\theta}_m\}+\epsilon_t(\bm s),
\end{equation}
where $Y_t(\bm{s})$ is the process at location $\bm{s}\in \mathcal{D}$ and time $t$, $\mathcal{M}\{\cdot; \boldsymbol{\theta}_m\}$ is an operator with parameters $\boldsymbol{\theta}_m$, $Y_{t-1}(\mathcal{D}) = \{Y_{t-1}(\bm r): \bm r\in\mathcal{D}\}$, and $\{\epsilon_{t}(\bm r): \bm r\in\mathcal{D}\}$ is a time-independent innovation process with zero mean, possibly spatially correlated and independent of $\{Y_{t'}(\mathcal{D}):t'<t\}$. This framework is quite general, accommodating both linearity and nonlinearity depending on $\mathcal{M}$ \citep{cressie}. Special cases include the vector autoregressive (VAR) model \citep[e.g.,][]{chris_space_time} and the integro-difference equation model \citep[e.g.,][]{wikle2002kernel,xu2005kernel}.

Despite its flexibility, direct application of \eqref{eqn:data_model} becomes computationally infeasible for large spatio-temporal datasets. A useful solution is low-rank approximation \citep{wikle1999dimension,wikle2001spatiotemporal,cressie2008fixed,annurev:/content/journals/10.1146/annurev-statistics-040120-020733}. Given a class of basis functions $\{\phi_k(\bm s):k=1,\ldots,\infty\}$ over $\mathcal{D}$, the process can be represented as $Y_t(\bm s)=\sum_{k=1}^\infty \alpha_{kt}\phi_k(\bm s)$, where $\{\alpha_{1t},\alpha_{2t}, \ldots\}$ are random expansion coefficients. Truncating the sum at $K>0$ and assuming $N$ locations, the process can be approximated as:
\begin{equation}
\label{eqn:data_model_discrete}
\bm{Y}_t=\bm{\Phi} \bm{\alpha}_t+\bm{\epsilon}_t,\quad \bm{\epsilon}_t \sim \text{Gau}(\bm{0},\sigma^2_d \bm{I}),
\end{equation}
where $\bm{Y}_t= ( Y_t(\bm{s}_1),...,Y_t(\bm{s}_N))^{\top}$, $\bm{\Phi}=[\phi_k(\bm s_i)]_{N \times K}$, $\bm{\alpha}_t = (\alpha_{1t},\ldots,\alpha_{Kt})^{\top}$, and $\bm{\epsilon}_t$ denotes measurement or approximation error. Such dimension reduction is often justified because the latent dynamics for most real-world spatio-temporal dynamic processes exist on a lower-dimensional manifold \citep[e.g., see the discussion in][]{wikle2019}.

The low-rank approximation shifts the focus to the dynamic evolution of the coefficient vector $\bm{\alpha}_t$, allowing \eqref{eqn:data_model} to be applied to $\bm{\alpha}_t$. For example, $\bm{\alpha}_t$ can be modelled using a first-order VAR model as:
\begin{equation}\label{eqn:VAR}
\boldsymbol{\alpha}_t=\bm{M} \boldsymbol{\alpha}_{t-1}+\bm{\omega}_t,
\end{equation}
where $\bm{M}\in\mathbb{R}^{K\times K}$ governs the linear evolution of the coefficients, and $\bm{\omega}_t=(\omega_{1t},\omega_{2t}, \ldots,\omega_{Kt})^{\top}$ represents innovations across the $K$ basis coefficients at time $t$. In practice, $\bm{\omega}_t$ is typically modelled as a multivariate Gaussian random vector as in the dynamic linear model framework \citep[e.g.,][]{west1989dynamic}, but this may be inadequate under extremal behaviour, as noted in Section \ref{sec:2}.  

\subsection{Accounting for extreme innovations}
\label{extreme_inno}
Heavier‑tailed models, such as stable or variance‑gamma distributions, better reflect extremal behaviour, especially in non‑stationary environmental settings. 
To utilize this flexibility while retaining interpretability, we place a latent regime‑switching structure on each innovation component $\omega_{kt}$. This process assumes that dynamics are governed by latent states, each determined by a hidden variable and its probability \citep[e.g., see][for time series]{Lange2009,Hamilton2010}. We model each state with a distribution for different tail behaviour, variance-gamma or stable for relatively heavy tails and Gaussian for lighter tails, as follows:
\begin{eqnarray}
\label{eqn:switch}
    \omega_{kt}|u_{kt}\sim 
    \begin{cases}
    u_{kt}\mathcal{S}(\lambda_{\text{S},k},\kappa_{\text{S},k},\nu_{\text{S},k},0)+(1-u_{kt})\text{Gau}(0,\sigma^2_{\text{Gau,}k}) \\ 
     u_{kt}\text{VG}(1,\theta_{\text{vg},k},\sigma_{\text{vg},k},0)+(1-u_{kt})\text{Gau}(0,\sigma^2_{\text{Gau,}k}), 
    \end{cases} 
\end{eqnarray}
where the latent variable satisfies $u_{kt} \sim \text{Bernoulli}(p_{kt})$ with $p_{kt}$ denoting its associated probability. The heavier-tailed distribution is used to capture extreme events through its slowly decaying tail property, while the light-tailed Gaussian distribution accounts for non-extreme events. This approach is both appealing and intuitive, as the distribution of innovations adapts to periods of anomalous volatility while maintaining the Gaussian benchmark during quieter phases. Furthermore, the estimated probability $p_{kt}=P(u_{kt}=1)$ represents the likelihood of an extreme event at time $t$ associated with $k^{\rm th}$ basis function. This framework defines a general class of DSTMs that represent heavy-tailed ($p_{kt}=1$), light-tailed ($p_{kt}=0$), or a mixture of both regimes ($ 0< p_{kt}<1$), thereby giving a probabilistic map of when and where extreme innovations are most likely to arise.

Beyond its intuitive interpretation, this regime-switching process also flexibly captures non-stationary extremal dependence, allowing smooth transitions between asymptotic dependence (AD) and independence (AI) in space and time (see Section \ref{sec:theo_res}). Crucially, whether AD or AI predominates can vary with spatial separation or temporal lag, so a model that adapts to these shifts is essential for accurately characterising tail behaviour.


\section{Theoretical Properties of the Model}\label{sec:theo_res}
In this section, we summarize the extremal dependence results under two distinct tail assumptions on the innovation vectors: regularly varying (RV) and exponentially-tailed (ET) innovations. We focus on the dependence properties of a bivariate random vector $(X_{1},X_{2})^{\top}$, as is common in the spatial extremes literature. The variables $X_1$ and $X_2$ correspond to spatial observations at different locations (and times). Specifically, we examine the measure $\chi_{12}(u)$, which quantifies the conditional probability that $X_2$ exceeds a threshold $u$ given that $X_1$ also exceeds that threshold \citep{huser2022advances}:
\begin{equation}\label{eqn:chi}
  \chi_{12}(u) = \Pr\{F_2(X_2) > u \mid F_1(X_1) > u\} = \frac{\Pr\{F_2(X_2) > u, F_1(X_1) > u\}}{\Pr\{F_2(X_2) > u\}},
\end{equation}
in which $u\in(0,1)$ and $F_1$ and $F_2$ are the continuous marginal distribution functions for $X_1$ and $X_2$, respectively. The range of $\chi_{12}(u)$ is between 0 and 1, where values closer to 1 indicate stronger extremal dependence between $X_1$ and $X_2$. When $u$ approaches 1, $\chi_{12}(u)$ evaluates the likelihood of one variable being extreme given that the other is similarly extreme. If $\chi_{12} = \lim_{u \rightarrow 1} \chi_{12}(u) = 0$, it suggests that $X_1$ and $X_2$ are \textit{asymptotically independent}, meaning that their asymptotic extreme behaviours are essentially unrelated. On the other hand, if $\chi_{12} = \lim_{u \rightarrow 1} \chi_{12}(u) > 0$, $X_1$ and $X_2$ are \textit{asymptotically dependent}, implying that their extreme values are dependent. This measure is valuable in understanding the joint extreme behaviour of spatially related variables (e.g., in studies of floods or heat waves). 

\subsection{General additive structure of the dynamic model}\label{sec:additive_structure}
To examine the extremal dependence properties of the combined models of \eqref{eqn:data_model} and \eqref{eqn:VAR}, we write the observed process at time $t$ as 
\begin{align}
\bm{Y}_t= \bm{\Phi}\bm{M}\bm{\alpha}_{t-1}+\bm{\Phi}\bm{\omega}_t+\bm{\epsilon}_t,
\end{align}
where the local spatial effects are determined by the innovation $\bm{\Phi}\bm{\omega}_t$, and the large scale (regional) effects are described by $\bm{\Phi}\bm{M}\bm{\alpha}_{t-1}$. Setting $\bm \alpha_0 = \bm\omega_0$ and iterating the state equation yields
$$
\bm \alpha_t = \sum_{l=0}^t \bm{M}^{t-l}\bm\omega_l,
$$
which is a linear transformation of the innovation vectors $\bm\omega_l$ at the time step $l$.

Substituting into the observation model, we obtain the representation:
\begin{equation}\label{eqn:linear_combination}
    \begin{split}
        \bm Y_t =& \sum_{l=0}^t \bm\Phi \bm{M}^{t-l}\bm\omega_l+\bm\varepsilon_t\equiv\bm\Psi_t\bm W_t+\bm\varepsilon_t,
    \end{split}
\end{equation}
in which $\bm\Psi_t = \bm\Phi(\bm M^{t}, \bm M^{t-1},\cdots,\bm M,\bm I)$, and $\bm W_t=(\bm\omega_0^{\top},\ldots, \bm\omega_t^{\top})^{\top}$.

Therefore, to study the upper-tail dependence properties, we focus on the linear transformation of the concatenated independent innovations $\bm W_t$. For notational conveniences (with slight abuse), we write $\bm W_t=(\bm\omega_0^{\top},\ldots, \bm\omega_t^{\top})^{\top}\equiv(w_1, \ldots, w_J)^{\top}$. Importantly, the formulation in~\eqref{eqn:linear_combination} facilitates the analysis of extremal dependence across both space and time. For any space-time pairs $(\bs_1, t_1)$ and $(\bs_2, t_2)$, let $X_1=Y_{t_1}(\bs_1)$ and $X_2=Y_{t_2}(\bs_2)$, which can be expressed as:
\begin{align}\label{eqn:additive_struc}
\begin{cases}
X_{1}= \psi_{11} w_{1}+\psi_{12} w_{2}+\cdots +\psi_{1J} w_{J},\\
X_{2}=\psi_{21} w_{1}+\psi_{22} w_{2}+\cdots+ \psi_{2J} w_{J},
\end{cases}
\end{align}
where $J=(\max\{t_1, t_2\}+1)K$. In the case when $t_1=t_2=t$, $(\psi_{i1}, \ldots, \psi_{iJ})^{\top}$ is simply equal to $(\bm\Psi_t)_{i,\cdot}$, the $i^{\rm th}$ row of $\bm\Psi_t$. 
The additive form in \eqref{eqn:additive_struc} therefore captures the complete bivariate dependence structure between any pair of spatial locations at time $t$, driven by the shared innovations $\bm W_t$ and modulated by the elements of ${\bm \Phi}$ and $\bm M$. 

For another example where $t_1=t$ and $t_2=t+1$, we examine how an extreme event at location $\bm s_1$ at time $t$ influences location $\bm s_2$ at time $t+1$. In this case, $J=(t+2)K$, and the corresponding coefficient vectors are
\begin{equation*}
    (\psi_{11}, \ldots, \psi_{1J})^{\top}=\{(\bm\Psi_t)_{1,\cdot},\bm 0^{\top}_K\}^{\top},\quad (\psi_{21}, \ldots, \psi_{2J})^{\top}=(\bm\Psi_{t+1})_{2,\cdot}
\end{equation*}
where $(\bm\Psi_t)_{i,\cdot}$ denote the $i^{\rm th}$ row vector of $\bm\Psi_t$, $i=1,2$.  This approach can be generalized to study the dependence between any space-time pair $(Y_{i_1, t_1},Y_{i_2, t_2})^{\top}$ for arbitrary $i_1\neq i_2$ and $t_1\neq t_2$.

In the following dependence property derivations, we condition on the vector $\bm u_l$, $l=1,\ldots, t$, i.e., we know the switching labels for being extreme or not extreme at each knot through time. By concatenating the labels across knots and times, we get $\bm U_t=(\bm u_1^{\top}, \ldots, \bm u_t^{\top})^{\top}\equiv (u_1,\ldots, u_J)^{\top}$. 

\subsection{Regularly varying innovations}
Consider the edge case $(u_1,\ldots, u_J)^{\top}=\bm 1_{J}$ and assume that all innovations across knots and times have independent, regularly varying components.  Given a non-negative matrix $\bm\Psi$ and $\bm X = \bm \Psi(w_1, \ldots, w_J)^{\top}$, 
we adapt results from \citet{Cooley2019} on the angular measure characterizing multivariate tail dependence in linear combinations of  independent regularly varying variables. We first consider the simplest case where the innovations have equal tail indices, but we extend the results in \citet{Cooley2019} to accommodate non-i.i.d. innovations and negative coefficients in $\bm \Psi$.

\begin{theorem}[Tail dependence for tail-equivalent stable innovations]
\label{theorem1}
    Consider the model as described in Section~\ref{sec:additive_structure}, condition on regime-switching labels $(u_1,\ldots, u_J)^{\top}=\bm 1_{J}$ in \eqref{eqn:switch}, and assume the innovations $(w_1, \ldots, w_J)^{\top}$ are independent, regularly varying components with stable distribution $ \mathcal{S}(\lambda_j,\kappa_j,\nu_j,\delta_j)$ and an equal concentration parameter $\lambda_j=\lambda$ for all $j\in \{1,\ldots, J\}$. For $(X_1,X_2)^{\top}$ constructed as in \eqref{eqn:additive_struc}, we can derive the asymptotic dependence coefficients for the upper-upper (UU), lower-lower (LL), and opposite-tail (UL, LU) dependencies. For $A\in\{UU,UL,LU,LL\}$, the dependence coefficients are given by:
    \begin{align*}
    \chi^A_{12}&=\sum_{J_A} \min\left\{\frac{|\psi_{1j}|^\lambda\nu_j^\lambda(1+\varphi^A_{1j}\kappa_j)}{\sum_{j=1}^J |\psi_{1j}|^\lambda\nu_j^\lambda(1+\varphi^A_{1j}\kappa_j)},\frac{|\psi_{2j}|^\lambda\nu_j^\lambda(1+\varphi^A_{2j}\kappa_j)}{\sum_{j=1}^J |\psi_{2j}|^\lambda\nu_j^\lambda(1+\varphi^A_{2j}\kappa_j)}\right\},
    \end{align*}
    where $J_A=\{j:\;\psi_{1j}\psi_{2j}>0\}$ if $A\in \{UU, LL\}$, $J_A=\{j:\;\psi_{1j}\psi_{2j}<0\}$ if $A\in \{UL, LU\}$, and 
    \begin{equation*}
        \begin{split}
            (\varphi^{UU}_{1j}, \varphi^{UU}_{2j})^{\top}=(\sign(\psi_{1j}),& \sign(\psi_{2j}))^{\top},\;(\varphi^{LL}_{1j}, \varphi^{LL}_{2j})^{\top}=(-\sign(\psi_{1j}), -\sign(\psi_{2j}))^{\top},\\
            (\varphi^{UL}_{1j}, \varphi^{UL}_{2j})^{\top}=(\sign(\psi_{1j}),& -\sign(\psi_{2j}))^{\top},\;(\varphi^{LU}_{1j}, \varphi^{LL}_{2j})^{\top}=(-\sign(\psi_{1j}), \sign(\psi_{2j}))^{\top}.
        \end{split}
    \end{equation*}
    Notably, opposite-tail dependence ($\chi^{\rm UL}$ and $\chi^{\rm LU}$) can only be nonzero when some coefficients $\psi_{ij}$ are negative.
\end{theorem}

Now we move away from the edge case where $(u_1,\ldots, u_J)^{\top}= \bm 1_{J}$ and allow the innovations at certain knots and times to be light-tailed. In this case, the summands in the expression of $(X_1, X_2)^{\top}$ in \eqref{eqn:additive_struc} have a joint tail  dominated by the terms with the smallest tail index (i.e., the most heavy-tailed ones), whereas the remaining terms do not contribute to the asymptotic extremal dependence properties. 

\begin{theorem}[Tail dependence for general stable innovations]
\label{theorem2}
    Suppose $(u_1,\ldots, u_J)^{\top}\neq \bm 1_{J}$ and the tail indices may differ even for innovations at $j\in\{j:u_j=1\}$. Let $\lambda_j=2$ for Gaussian innovations, and denote $\tilde{\lambda}=\min_{j=1,\ldots, J} \lambda_j$ as the smallest tail index. Define 
$$
\mathcal{T} = \{j: \max_{i=1,2}\psi_{ij}> 0,\, \lambda_j = \tilde{\lambda}\},
$$ 
as the set of indices contributing the heaviest tails. When $\mathcal{T}\neq \emptyset$, the extremal dependence coefficients are determined exclusively by these innovations, while lighter-tailed terms become negligible. Consequently, the extremal dependence results derived under equal-tail conditions apply directly to the subset $\mathcal{T}$.
\end{theorem}


\subsection{Exponentially-tailed innovations}

Now we assume innovations $\omega_j$ belong to $\mathrm{ET}^\infty_{\lambda_j^+}$ (upper exponential tail with rate parameter $\lambda_j^+$) and $\mathrm{ET}^{-\infty}_{\lambda_j^-}$ (lower exponential tail with rate parameter $\lambda_j^-$). The following result specifically covers variance-gamma innovations.

\begin{theorem}[Tail dependence for exponential-tailed innovations]\label{thm:ET}
    Assume that the independent variables $\omega_j$ are in $\mathrm{ET}^\infty_{\lambda_j^+}$  and $\mathrm{ET}^{-\infty}_{\lambda_j^-}$, where we use the convention that a variable with tail lighter than exponential has $\lambda^\pm_j=\infty$. Define scale parameters associated with dependence directions:
$$
\bm{\psi}^{A}_j=\frac{1}{\lambda_j^+}\max\{\bm{\psi}^{A}_j,\bm{0}\}+\frac{1}{\lambda_j^-}\max\{-\bm{\psi}^{A}_j,\bm{0}\}, \quad A\in\{UU,UL,LU,LL\},
$$
where $\bm{\psi}^{UU}_j=(\psi_{1j},\psi_{2j})^{\top}$, $\bm{\psi}^{UL}_j=(\psi_{1j},-\psi_{2j})^{\top}$, $\bm{\psi}^{LU}_j=(-\psi_{1j},\psi_{2j})^{\top}$, and $\bm{\psi}^{LL}_j=(-\psi_{1j},-\psi_{2j})^{\top}$. Set $\tilde{\psi}_i^{A}=\max_j \psi_{ij}^{A}$ and let $I_i^{A}$ be the set of indices where this maximum occurs. If $I_1^{A}=I_2^{A}$ and $\tilde{\psi}_i^{A}>0$ for $i=1,2$, then $Y_1$ and $Y_2$ exhibit asymptotic dependence with
$$
\chi^{A}_{12}=\mathbb{E}\left[\min\left\{\frac{\exp(\tilde{X}_1^{A})}{m_1^{A}}, \frac{\exp(\tilde{X}_2^{A})}{m_2^{A}}\right\}\right]>0,
$$
where 
$$
(\tilde{X}_1^{A}, \tilde{X}_2^{A})^{\top}=\sum_{j\not\in I_1^{A}} \omega_{j}\left(\frac{\psi_{1j}^A}{\tilde{\psi}_{1}^{A}}, \frac{\psi_{2j}^A}{\tilde{\psi}_{2}^{A}}\right)^{\top},\quad m_i^{A}=\mathbb{E}[\exp(\tilde{X}_i^{A})],\quad i=1,2.
$$
Thus, asymptotic dependence arises when the same subset of innovations dominates both components' tails.
\end{theorem}

\subsection{Long-term extremal dependence}
 \citet{cressie} emphasized the importance of the dynamic system corresponding to a DSTM being \textit{stable}, i.e., the VAR process should oscillate around a finite long‑run mean rather than grow without bound. This is satisfied when all eigenvalues of the transition matrix $\bm M$ lie strictly inside the unit circle. Equivalently, its spectral radius $\|\bm{M}\|=\max\{|\lambda_i|:i=1,\ldots, K\}<1$, in which $\{\lambda_1, \ldots, \lambda_{K}\}$ are all the eigenvalues of $\bm M$. 

In addition, we want the transition matrix $\bm M$ to be non‑defective (or diagonalizable), i.e., its algebraic and geometric multiplicities coincide for every eigenvalue. This assumption adds an extra layer of mathematical and numerical security. Under the stable and non‑defective assumptions, the VAR system converges to a non‑explosive equilibrium with finite mean and covariance. 

Beyond the first and second-order moments, we want to examine whether the extremal dependence also approaches a steady value as $t\rightarrow \infty$. We investigate this question in what follows.
\begin{theorem}\label{thm:long_term}
    Assume that the transition matrix $\bm{M}\in \mathbb{R}^{K\times K}$ is non-defective, and its spectral radius $\|\bm{M}\|<1$. Then for any fixed $t\in\mathbb{N}^+$, any two components from $\bm Y_{t}$ and $\bm Y_{t+\Delta t}$, respectively, tend towards asymptotic independence as the time gap increases. That is, any pair has a dependence measure that satisfies $\chi^A\rightarrow 0$ as $\Delta t\rightarrow\infty$ for $A\in \{UU, LL, LU, UL\}$.
\end{theorem}
\noindent The assumptions in Theorem~\ref{thm:long_term} are standard practice in multivariate time series analysis of VAR models to ensure consistent estimators, ergodicity, stationarity, and practical utility in forecasting \citep[e.g.,][]{lutkepohl2013introduction}. In our context, this facilitates the derivation of long-term tail-dependence coefficients as $\Delta t$ approaches infinity, since these coefficients depend on how rapidly the innovations at earlier times are damped by $\bm M^{\Delta t}$.


\subsection{Examples}\label{sec:example}
\begin{figure}[!t]
\centering
\includegraphics[width=0.78\linewidth]{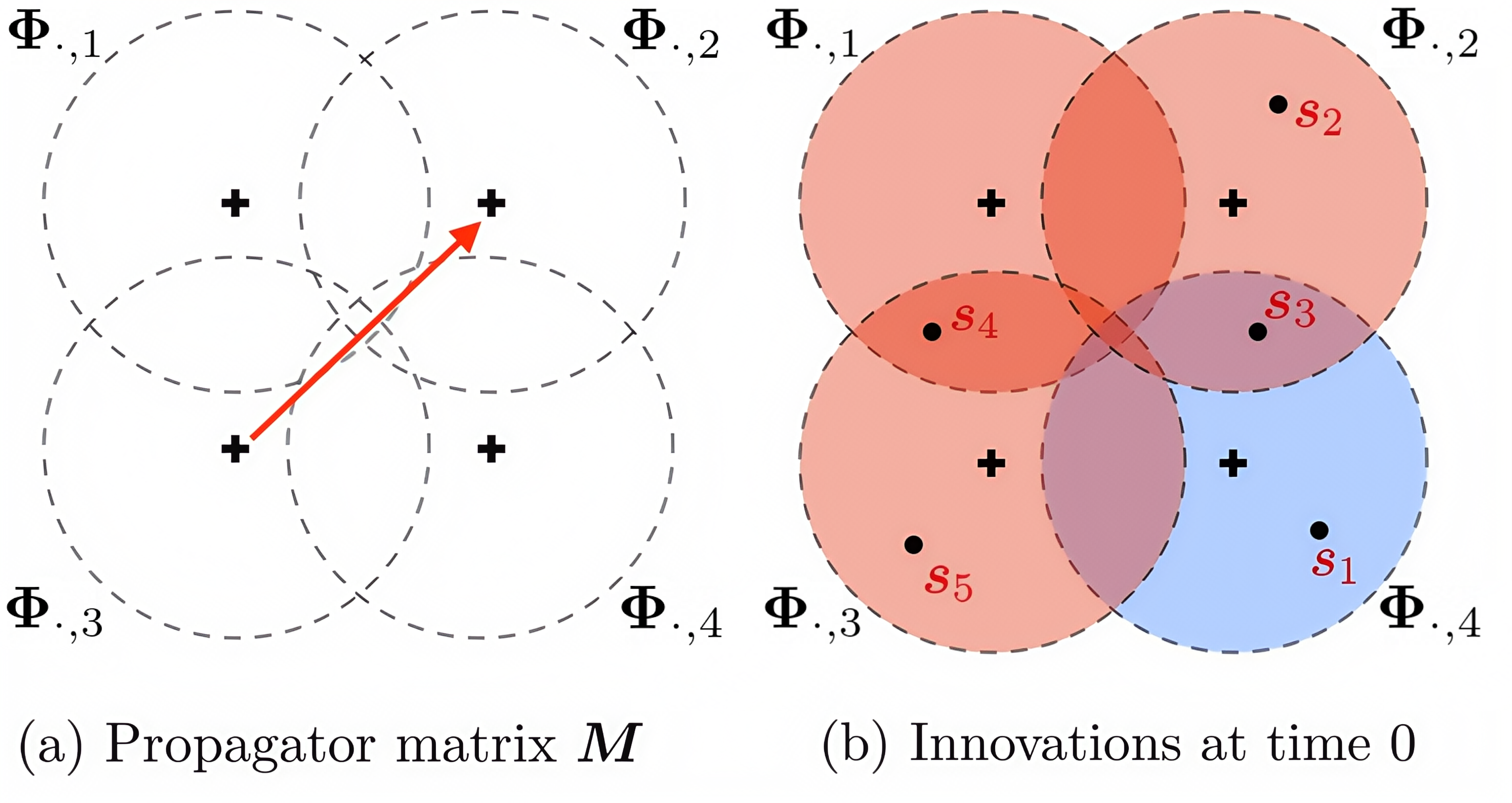} 
\vskip -0.3cm
  \caption{Illustration of Theorem~\ref{theorem1} with $K=4$. An extreme event occurs within regions marked in red, while the region in blue does not observe extreme events. Black cross marks indicate the knot points, and black dots indicate locations. Each region is covered by one of the local basis functions $\bm\Phi_{\cdot,1},\ldots,\bm\Phi_{\cdot,4}$.}
    \label{fig1_theory}
\end{figure}

To further investigate the expression of the tail dependence coefficient $\chi$ in Theorem~\ref{theorem1}, we
first consider the simple example over three time points, $t=0,\;1,\;2$, where the spatial process is described by $K=4$ local Wendland basis functions denoted by $\bm\Phi_{\cdot,1},\ldots,\bm\Phi_{\cdot,4}$; see Figure~\ref{fig1_theory}(b). Then $\bm M\in \mathbb{R}^{4\times 4}$, and we assume that its diagonal elements satisfy $m_{kk}\in (0,1)$, $k=1,\ldots, 4$ (i.e., the process will diminish over time locally) and only one off-diagonal element is nonzero, $m_{23}\in (0,1)$, i.e., the extremes around knot 3 can be moved to the region covered by knot 2 from time $t$ to $t+1$;  see Figure~\ref{fig1_theory}(a). Meanwhile, for locations $\bm s_1,\ldots,\bm s_5$ marked in Figure~\ref{fig1_theory}(b), we have $\bm \Phi\in \mathbb{R}^{5\times 4}$ (row-standardized). Locations $\bm s_1$, $\bm s_2$ and $\bm s_5$ are only covered by one knot, and therefore their corresponding rows of $\bm \Phi$ only have one nonzero element $1$ arising from the corresponding local basis function. Locations $\bm s_3$ and $\bm s_4$ were placed at equal distance to the  two knots covering them, and therefore $\bm \Phi_{3, \cdot} = (0,1/2,0, 1/2)$ and $\bm \Phi_{4, \cdot} = (1/2,0, 1/2,0)$.

First, we assume at time $t=0$, $\bm u_0=(1,1,1,0)^{\top}$, i.e., innovations are stable distributed at the first three knots but normally distributed at the fourth knot, i.e., $\omega_{k0}\sim \mathcal{S}(\lambda, \kappa_k, \nu_k, \delta_k)$ with $\lambda\in (0,2)$, $k=1,\;2,\;3$, and $\omega_{40}\sim \text{Gau}(0,\sigma^2_4)$. At times $t=1$ and $t=2$, we assume $\bm u_1=\bm u_2=\bm 0$, i.e., the innovations at the latter two times are not heavy-tailed. Effectively, we assume there is a storm that emerges in the upper-left of the spatial domain at the initial time, and we illustrate how the extremal dependence evolves over space and time. 

By the reasoning in Section~\ref{sec:additive_structure}, we can vectorize all the innovations from these three times, and write $\bm Y_t = \bm\Psi_t\bm W_t+\bm\varepsilon_t$, where $\bm\Psi_t = \bm\Phi(\bm M^{t}, \bm M^{t-1},\cdots,\bm M,\bm I)$, and $\bm W_t=(\bm\omega_0^{\top},\ldots, \bm\omega_t^{\top})^{\top}$, $t=0,\;1,\;2$. The specific forms of the expanded $\bm\Psi_t$ are given in Section~\ref{sec:example_proof} of the Supplementary Material. Then by Theorems~\ref{theorem1} and \ref{theorem2}, we can easily derive the $\chi$-dependence measure between any two locations at any times, which are listed in Table~\ref{tab:chi_example1}.
\begin{table}[]
\caption{The $\chi$-dependence coefficients for the configuration depicted in Figure \ref{fig1_theory}, where extreme innovations occur only at knots $1$--$3$ at the initial time point $t=0$. We report the upper–upper ($\chi^{UU}$) and lower–lower ($\chi^{LL}$) coefficients; the opposite-tail coefficients $\chi^{LU}$ and $\chi^{UL}$ are identically zero in this setting. Although the listed $\chi$ values may appear independent of $\bm\Phi$, they do in fact depend on it---the apparent invariance results from fixing specific values for the elements of $\bm\Phi$. See Section \ref{sec:example_proof} of the Supplementary Material for a detailed derivation.}
\label{tab:chi_example1}

{\renewcommand{\arraystretch}{2.8}
\resizebox{\textwidth}{!}{%
\begin{tabular}{c|cc}
\Xhline{3\arrayrulewidth}
                            & $\chi^{UU}$                                                                                                                                                                                                                                                                                         & $\chi^{LL}$                                                                                                                                                                                                                                                                                         \\ \Xhline{3\arrayrulewidth}
$(Y_{\bm s_4,0},Y_{\bm s_5,0})^{\top}$ & $\frac{\nu_3^{\lambda}(1+\kappa_3)}{\nu_1^{\lambda}(1+\kappa_1)+\nu_3^{\lambda}(1+\kappa_3)}$                                                                                                                                                                                        & $\frac{\nu_3^{\lambda}(1-\kappa_3)}{\nu_1^{\lambda}(1-\kappa_1)+\nu_3^{\lambda}(1-\kappa_3)}$                                                                                                                                                                                        \\ \hline
$(Y_{\bm s_4,1},Y_{\bm s_5,1})^{\top}$ & $\frac{(m_{33}\nu_3)^{\lambda}(1+\kappa_3)}{(m_{11}\nu_1)^{\lambda}(1+\kappa_1)+(m_{33}\nu_3)^{\lambda}(1+\kappa_3)}$                                                                                                                                                                & $\frac{(m_{33}\nu_3)^{\lambda}(1-\kappa_3)}{(m_{11}\nu_1)^{\lambda}(1-\kappa_1)+(m_{33}\nu_3)^{\lambda}(1-\kappa_3)}$                                                                                                                                                                \\ \hline
$(Y_{\bm s_4,2},Y_{\bm s_5,2})^{\top}$ & $\frac{(m_{33}^2\nu_3)^{\lambda}(1+\kappa_3)}{(m_{11}^2\nu_1)^{\lambda}(1+\kappa_1)+(m_{33}^2\nu_3)^{\lambda}(1+\kappa_3)}$                                                                                                                                                          & $\frac{(m_{33}^2\nu_3)^{\lambda}(1-\kappa_3)}{(m_{11}^2\nu_1)^{\lambda}(1-\kappa_1)+(m_{33}^2\nu_3)^{\lambda}(1-\kappa_3)}$                                                                                                                                                          \\ \hline
$(Y_{\bm s_4,0},Y_{\bm s_2,0})^{\top}$ & 0                                                                                                                                                                                                                                                                                                   & 0                                                                                                                                                                                                                                                                                                   \\ \hline
$(Y_{\bm s_4,0},Y_{\bm s_2,1})^{\top}$ & $\min\left\{\frac{\nu_3^{\lambda}(1+\kappa_3)}{\nu_1^{\lambda}(1+\kappa_1)+\nu_3^{\lambda}(1+\kappa_3)}, \frac{(m_{23}\nu_3)^{\lambda}(1+\kappa_3)}{(m_{22}\nu_2)^{\lambda}(1+\kappa_2)+(m_{23}\nu_3)^{\lambda}(1+\kappa_3)}\right\}$                                 & $\min\left\{\frac{\nu_3^{\lambda}(1-\kappa_3)}{\nu_1^{\lambda}(1-\kappa_1)+\nu_3^{\lambda}(1-\kappa_3)}, \frac{(m_{23}\nu_3)^{\lambda}(1-\kappa_3)}{(m_{22}\nu_2)^{\lambda}(1-\kappa_2)+(m_{23}\nu_3)^{\lambda}(1-\kappa_3)}\right\}$                                 \\ \hline
$(Y_{\bm s_4,0},Y_{\bm s_2,2})^{\top}$ & {\footnotesize $\min\left\{\frac{\nu_3^{\lambda}(1+\kappa_3)}{\nu_1^{\lambda}(1+\kappa_1)+\nu_3^{\lambda}(1+\kappa_3)}, \frac{(m_{23}(m_{22}+m_{33})\nu_3)^{\lambda}(1+\kappa_3)}{(m_{22}^2\nu_2)^{\lambda}(1+\kappa_2)+(m_{23}(m_{22}+m_{33})\nu_3)^{\lambda}(1+\kappa_3)}\right\}$} & {\footnotesize $\min\left\{\frac{\nu_3^{\lambda}(1-\kappa_3)}{\nu_1^{\lambda}(1-\kappa_1)+\nu_3^{\lambda}(1-\kappa_3)}, \frac{(m_{23}(m_{22}+m_{33})\nu_3)^{\lambda}(1-\kappa_3)}{(m_{22}^2\nu_2)^{\lambda}(1-\kappa_2)+(m_{23}(m_{22}+m_{33})\nu_3)^{\lambda}(1-\kappa_3)}\right\}$} \\ \hline
{\renewcommand{\arraystretch}{1} \begin{tabular}[c]{@{}c@{}}$(Y_{\bm s_1,t},Y_{\bm s_i,t})^{\top}$\\ $i\neq 1$, $t\geq 0$\end{tabular}} &0&0\\ \hline
{\renewcommand{\arraystretch}{1} \begin{tabular}[c]{@{}c@{}}$(Y_{\bm s_2,t},Y_{\bm s_2,t+1})^{\top}$\\ $t\geq 0$\end{tabular}} &$\frac{(m_{23}\nu_3\sum_{t_1=0}^tm_{22}^{t_1}m_{33}^{t-t_1})^{\lambda}(1+\kappa_3)}{(m_{22}^{t+1}\nu_2)^{\lambda}(1+\kappa_2)+(m_{23}\nu_3\sum_{t_1=0}^tm_{22}^{t_1}m_{33}^{t-t_1})^{\lambda}(1+\kappa_3)}$&$\frac{(m_{23}\nu_3\sum_{t_1=0}^tm_{22}^{t_1}m_{33}^{t-t_1})^{\lambda}(1-\kappa_3)}{(m_{22}^{t+1}\nu_2)^{\lambda}(1-\kappa_2)+(m_{23}\nu_3\sum_{t_1=0}^tm_{22}^{t_1}m_{33}^{t-t_1})^{\lambda}(1-\kappa_3)}$\\\Xhline{3\arrayrulewidth}
\end{tabular}}
}
\end{table}

Locations $\bm s_4$ and $\bm s_5$ are both covered by the local basis function centred at the third knot which experiences a stable innovation at the initial time. Therefore, they exhibit AD at the upper and lower tail at time $0$. Since $\bm s_4$ is also affected by the first knot, we see that the dependence strength also hinges on the skewness $\kappa_1$ and the scale $\nu_1$ of the innovation at knot 1 and time 0, i.e., $\omega_{10}$. On the other hand, since $0<m_{33}<1$, the extremes stemming from $\omega_{30}$ will slowly decay over time, and thus the $\chi$-coefficient decreases monotonically from  $(Y_{\bm s_4,0},Y_{\bm s_5,0})^{\top}$ to $(Y_{\bm s_4,2},Y_{\bm s_5,2})^{\top}$. Furthermore, we show in Section~\ref{sec:example_proof} of the Supplementary Material that 
\begin{equation}\label{eqn:RV_example}
    \chi(Y_{\bm s_4,t},Y_{\bm s_5,t})=\frac{(m_{33}^t\nu_3)^{\lambda}(1+\kappa_3)}{(m_{11}^t\nu_1)^{\lambda}(1+\kappa_1)+(m_{33}^t\nu_3)^{\lambda}(1+\kappa_3)}, \forall\; t\in \mathbb{N}^+,
\end{equation}
and $\chi(Y_{\bm s_4,t},Y_{\bm s_5,t})\rightarrow 0$ as $t\rightarrow \infty$, which is consistent with the results on long-term dependence in Theorem~\ref{thm:long_term}.

In comparison, locations $\bm s_4$ and $\bm s_2$ do not share any local basis functions and thus exhibit AI at time $0$. However, the propagator $\bm M$ allows the extremes to transition from the third knot to the second knot ($m_{23}>0$; see Figure~\ref{fig1_theory}). As a result, we see that $\chi(Y_{\bm s_4,0},Y_{\bm s_2,t})> 0$ for $t=1,2$ and the extremes stemming from $\omega_{30}$ will propagate to $\bm s_2$ at later times even though all later innovations are Gaussian. Similarly to the previous case, we can also show that $\chi(Y_{\bm s_4,0},Y_{\bm s_2,t})\rightarrow 0$ as $t\rightarrow \infty$.

Interestingly, location $\bm s_1$ is only covered by the fourth basis function, which is not subject to heavy-tailed innovations at any time. In addition, the transition matrix does not transfer the extremes from the other knots with extremes. Therefore, we have $\chi(Y_{\bm s_1,t},Y_{\bm s_i,t})= 0$ for $i=2,\ldots,5$ and any $t\geq 0$.

Lastly, for any finite $t\geq 0$, $\chi(Y_{\bm s_2,t},Y_{\bm s_2,t+1})>0$ for $t\geq 0$, demonstrating persistent local dependence. In this respect, notable differences would arise with exponential-tailed innovations. Using local basis functions,  the dominating innovations determining the presence of asymptotic dependence in Theorem~\ref{thm:ET} will (in practice) typically be different for two locations near different knots and/or at different time steps. Then, AD arises only locally in space around a knot for a given time, whereas we observe AI otherwise. Therefore, while AD will typically propagate through space and time according to the dynamics specified by $\bm M$ with stable (i.e., regularly varying) innovations, it is a property arising only locally in space and time with exponential-tailed models.

Even for this simple space-time dynamic setup, our regime-switching allows a rich spectrum of dependence properties at different locations and times. Crucially, our model also captures the temporal extremal dependence. To the best of our knowledge, local extremal dependence that varies over time has received little attention in the spatial extremes literature, and no existing method adequately captures such dependence. Also, our setup is realistic and interpretable, since the transition matrix $\bm M$ heavily governs the dependence structure. Although the $\chi$-coefficients in Table~\ref{tab:chi_example1} seem to have complicated expressions, it is easy to use Theorems~\ref{theorem1} and \ref{theorem2} to calculate closed forms after writing any space-time pairs into the additive structure via calculating the specific forms of $\bm \Psi_t$ for time 0 to a maximum time of interest.

Lastly, the result of Theorem \ref{theorem1} also applies for global basis functions which may lead to a basis function matrix $\bm\Phi$ with negative elements, and hence also negative elements for $\bm\Psi_t$, $t\in \mathbb{N}^+$ (so we have non-zero $\chi^{LU}$ and $\chi^{UL}$ in certain cases). In Section~\ref{sec:example_ET} of the Supplementary Material, we illustrate this with exponentially-tailed innovations and global thin‑plate bases, where we indeed observe non‑zero dependence in all four quadrants $(UU,LL,UL,LU)$, and especially non-zero opposite tail dependence. Several existing spatial models already allow limited differentiation between upper‑ and lower‑tail behaviour, for example the separable‑tail construction of \citet{gong2022flexible}, or elliptical copula models with symmetry between UU and LL, and between UL and LU, and the same dependence regime (asymptotic dependence or independence) in all four quadrants \citep{HUSER2017166}. \textit{What distinguishes the present framework is that the sign‑flexible basis yields simultaneous, non‑separable control over upper–upper, lower–lower, and the two opposite‑tail quadrants within a single spatio‑temporal process.} This offers a partial answer to the open problem raised in \citet{krock2023tail} who advocate to build such a spatial modelling framework.

Similar results can be shown for the other two cases: (1) regularly varying innovations and global thin-plate basis functions and (2) exponentially-tailed innovations and local basis functions, although these are omitted for brevity.


\section{Bayesian inference}\label{sec:Bayes_inference}

The extreme dynamic spatio-temporal model in Section \ref{sec:3} fits naturally within the Bayesian hierarchical framework, which is widely used for dependent data such as dynamic linear models and DSTMs \citep[e.g.,][]{west1989dynamic, cressie}. It comprises data, process, and parameter models (Sections \ref{sec:3.2}), with inference carried out using a Bayesian algorithm (Section \ref{sec:3.4}).

\subsection{Bayesian hierarchical model}
\label{sec:3.2} 
Our BHM begins with the data model in~\eqref{eqn:data_model_discrete}. Then, the process model outlined in~\eqref{eqn:VAR}--\eqref{eqn:switch} describes the dynamics of the coefficients $\bm{\alpha}_t$ and the regime-switching process for $\bm{\omega}_t$ through the indicator $u_{kt}\sim \text{Bernoulli}(p_{kt})$. To let the switching probability vary smoothly in time, we express it with a logistic spline:
\begin{eqnarray*}
 p_{kt}&=& \frac{1}{1+\exp(-1 \cdot \bm{B}(t)^{\top}\bm{\beta}_k)}, 
\end{eqnarray*}
where $\bm{B}(t)$ is a vector of temporal basis functions and $\bm{\beta}_k$ is the corresponding coefficient vector for mode $k$. In our applications, $\bm{B}(t)$ consists of cubic B‑splines, which give a parsimonious representation of the time‑varying regime probabilities.

In \eqref{eqn:VAR}, parameterization of the transition matrix is critical, as the structure of $\bm{M}$ determines the association between $\bm{\alpha}_{t-1}$ and $\bm{\alpha}_t$ \citep{wikle2019}. In our real data analysis, we utilize a lagged-nearest neighbor structure \citep{chris_space_time}, with $\bm{\Phi}$ constructed from Wendland local basis functions, whose knots $\bm{s}_k^* = ( s_{k,1}^*, s_{k,2}^*)^{\top}$ for $k = 1, \ldots, K$ are placed uniformly over $\mathcal{D}$. This structure induces dependence only among neighboring knots, ensuring sparsity in $\bm{M}$. Then, \eqref{eqn:VAR} becomes $\alpha_{s^*_{k,1},s^*_{k,2}}(t)= b({s^*_{k,1},s^*_{k,2}}) \cdot \alpha_{s^*_{k,1},s^*_{k,2}}(t-1)  + c \cdot \alpha_{s^*_{k+1,1},s^*_{k,2}}(t-1) + d \cdot \alpha_{s^*_{k,1},s^*_{k+1,2}}(t-1) + e \cdot \alpha_{s^*_{k-1,1},s^*_{k,2}}(t-1)+ f \cdot \alpha_{s^*_{k,1},s^*_{k-1,2}}(t-1) +\omega_{s^*_{k,1},s^*_{k,2}}(t)$, where $b({s^*_{k,1},s^*_{k,2}})$, $c$, $d$, $e$ and $f$ represent the nearest neighbor parameters, respectively. Note that $b({s^*_{k,1},s^*_{k,2}})$ varies spatially while $c$, $d$, $e$, and $f$ are constant across space. 

We complete the BHM by assigning prior distributions to every parameter in the data and process layers; details appear in Section~\ref{sec:prior_model} of the Supplementary Material.

\subsection{MCMC Sampling Algorithm}
\label{sec:3.4}
Denoting the set of all parameters by $\bm{\Theta}$, Bayesian inference is based on the posterior distribution of $\bm{\Theta}$: $\pi(\bm{\Theta}|-)\propto \pi(\bm{y}_1,...,\bm{y}_T|\bm{\Theta}) \cdot \pi(\bm{\Theta})$, where $\pi(\bm{y}_1,...,\bm{y}_T|\bm{\Theta})$ and $\pi(\bm{\Theta})$ represent the likelihood and prior distribution, respectively. Nearly all parameters in our model require updating using the Metropolis-Hastings algorithm \citep{fa181ebf-f6d7-3695-81c0-38bc48246232} due to the lack of conjugacy (variance-gamma distribution) and the lack of a closed-form density function (stable distribution). Although feasible for moderate $K$ and $T$, the method requires careful tuning and may become computationally prohibitive as $K$ and $T$ grow.  To address this, we introduce a nuisance term $\bm{\eta}_t$ in the process model:
\begin{equation}
\label{eqn:4.3.12}
\bm{\alpha}_t =\bm{M} \bm{\alpha}_{t-1}+\bm{\omega}_t +\bm{\eta}_t, \quad \bm{\eta}_t=[\eta_{1t},...,\eta_{Kt}]^{\top}, \quad \bm{\eta}_t\sim Gau(\bm{0},\sigma^2_{\eta}\bm{I}),
\end{equation}
where $\sigma^2_{\eta}$ is a fixed, small positive constant controlling the scale of $\bm{\eta}_t$ and serving as a tuning parameter. Given $\bm{\omega}_t$, both $\bm{M}$ and $\bm{\alpha}_t$ can then be updated via Gibbs sampling. When $\sigma^2_{\eta}$ is small, the effect of $\bm{\eta}_t$ on $\bm{\alpha}_t$ is negligible. We examine its impact in Section \ref{sec:5}.
\section{Simulation experiments}\label{sec:5}
This section presents two simulation studies: one demonstrating that the MCMC algorithm with the nuisance term in \eqref{eqn:4.3.12} recovers true parameters, and another showing that the model detects extreme events under model mis-specification.
\subsection{Parameter estimation with the nuisance term}
\begin{figure}[t]
\centering
\includegraphics[width=0.9\linewidth]{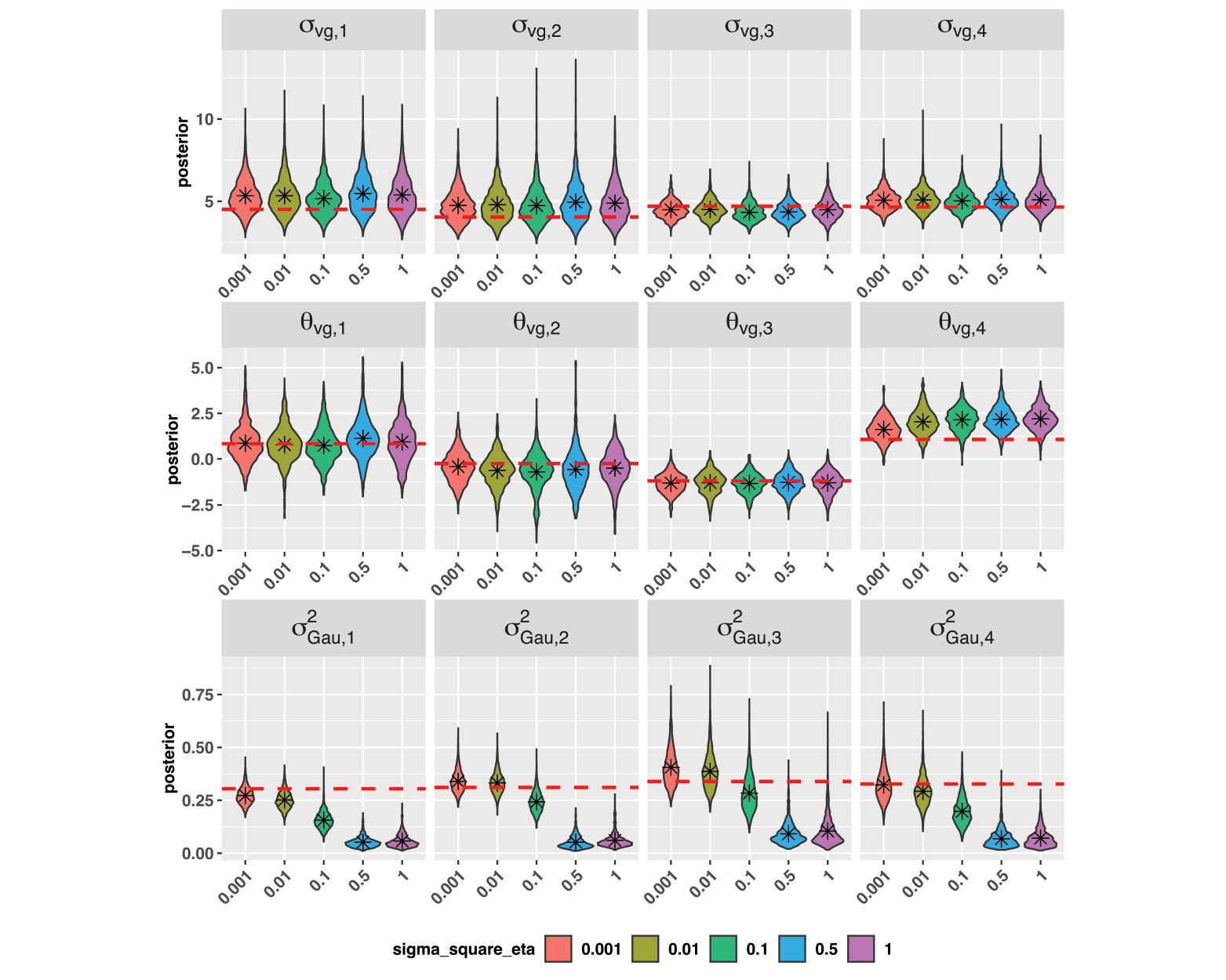}
  \caption{Violin plots of $\sigma_{\text{vg},k}$, $\theta_{\text{vg},k}$, and $\sigma^2_{\text{Gau},k}$ for $k = 1, \dots, 4$ (from left to right), corresponding to different values of $\sigma^2_{\eta}$. The red dashed horizontal lines represent the true values.}
  \label{variance_gamma_estimate}
\end{figure}
\begin{figure}[!t]
\centering
\includegraphics[width=1\linewidth]{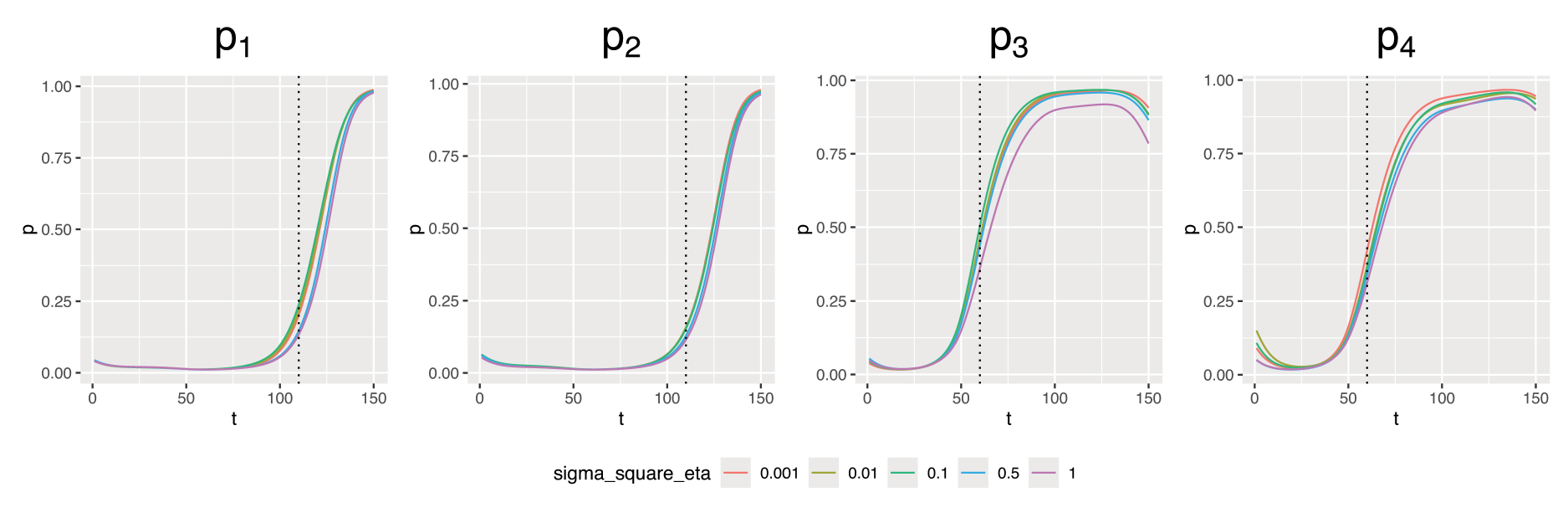}
  \caption{Posterior mean of $p_{kt}$ for $k=1,...,4$ (from left to right) and $t=1,...,T$ with varying $\sigma^2_{\eta}$.}
    \label{simul_fig3}
\end{figure}
We consider a spatio-temporal process with $N=900$ spatial locations and $T=150$ time points, and first generate the innovation $\omega_{kt}$ by:
\begin{equation}
\label{simul_generate}
\pi(\omega_{kt}) = 
\begin{cases}
\pi\big(\text{Gau}(0, \sigma^2_{\text{Gau},k})\big), & k \in \{1,2\}, \ t \leq 120 \\[2pt]
\pi\big(\text{VG}(1,\theta_{\text{vg},k},\sigma_{\text{vg},k},0)\big), & k \in \{1,2\}, \ t > 120 \\[2pt]
\pi\big(\text{Gau}(0, \sigma^2_{\text{Gau},k})\big), & k \in \{3,4\}, \ t \leq 60 \\[2pt]
\pi\big(\text{VG}(1,\theta_{\text{vg},k},\sigma_{\text{vg},k},0)\big), & k \in \{3,4\}, \ t > 60,
\end{cases}
\end{equation}

where we use the variance-gamma distribution to represent the heavier-tailed component. We generate $\bm{\alpha}_t$ using \eqref{eqn:VAR}, assuming a ``rook"-type 1-lagged nearest-neighbour structure for $\bm{M}$. We construct $\bm{\Phi}$ using local Gaussian radial basis functions with $K = 4$ uniformly spaced knots and generate data at $N$ locations and $T$ time points via \eqref{eqn:data_model_discrete}. By construction, $\omega_{kt}$ becomes exponential-tailed for $k=1,2$ after $t>120$ and for $k=3,4$ after $t>60$, so $p_{kt}$ in the switching process is expected to increase accordingly.

We fit the model using $q=5$ temporal
B-spline basis functions $\bm{B}$ with a piecewise polynomial of degree 2, using a weakly informative prior for $\sigma_{\text{vg},k}$ ($r_{\sigma}=2$, $\gamma_{\sigma}=4.5$) and vague priors for other parameters. We draw 20,000 posterior samples, discarding the first 10,000 as burn-in. Figure \ref{variance_gamma_estimate} shows violin plots of $\sigma_{\text{vg},k}$, $\theta_{\text{vg},k}$, and $\sigma^2_{\text{Gau},k}$ for varying $\sigma^2_{\eta}=(0.001,0.01,0.1,0.5,1)$, along with the true parameters. Estimates of $\sigma_{\text{vg},k}$ and $\theta_{\text{vg},k}$ remain accurate, while $\sigma^2_{\text{Gau},k}$ grows increasingly biased as $\sigma^2_{\eta}$ increases, indicating the greater influence of the nuisance term $\bm{\eta}_t$ on $\bm{\alpha}_t$, which can make parameter estimation challenging for $\bm{\omega}_t$.

Figure \ref{simul_fig3} shows the effect of $\bm{\eta}_t$ on estimating $\bm{p}_t$. Despite underestimating $\sigma^2_{\text{Gau},k}$ for large $\sigma^2_{\eta}$, posterior means of $\bm{p}_t$ remain consistent, and align well with the data generation in \eqref{simul_generate}. Specifically, $p_{1,t}$ and $p_{2,t}$ rise after $t=120$, and $p_{3,t}$ and $p_{4,t}$ after $t=60$, demonstrating that the true parameters can be estimated. A similar analysis for the stable distribution is in Supplementary Material \ref{sec:stable_estimate}.

\subsection{Detection of extreme events}
\begin{figure}[!t]
\centering
\includegraphics[width=1\linewidth]{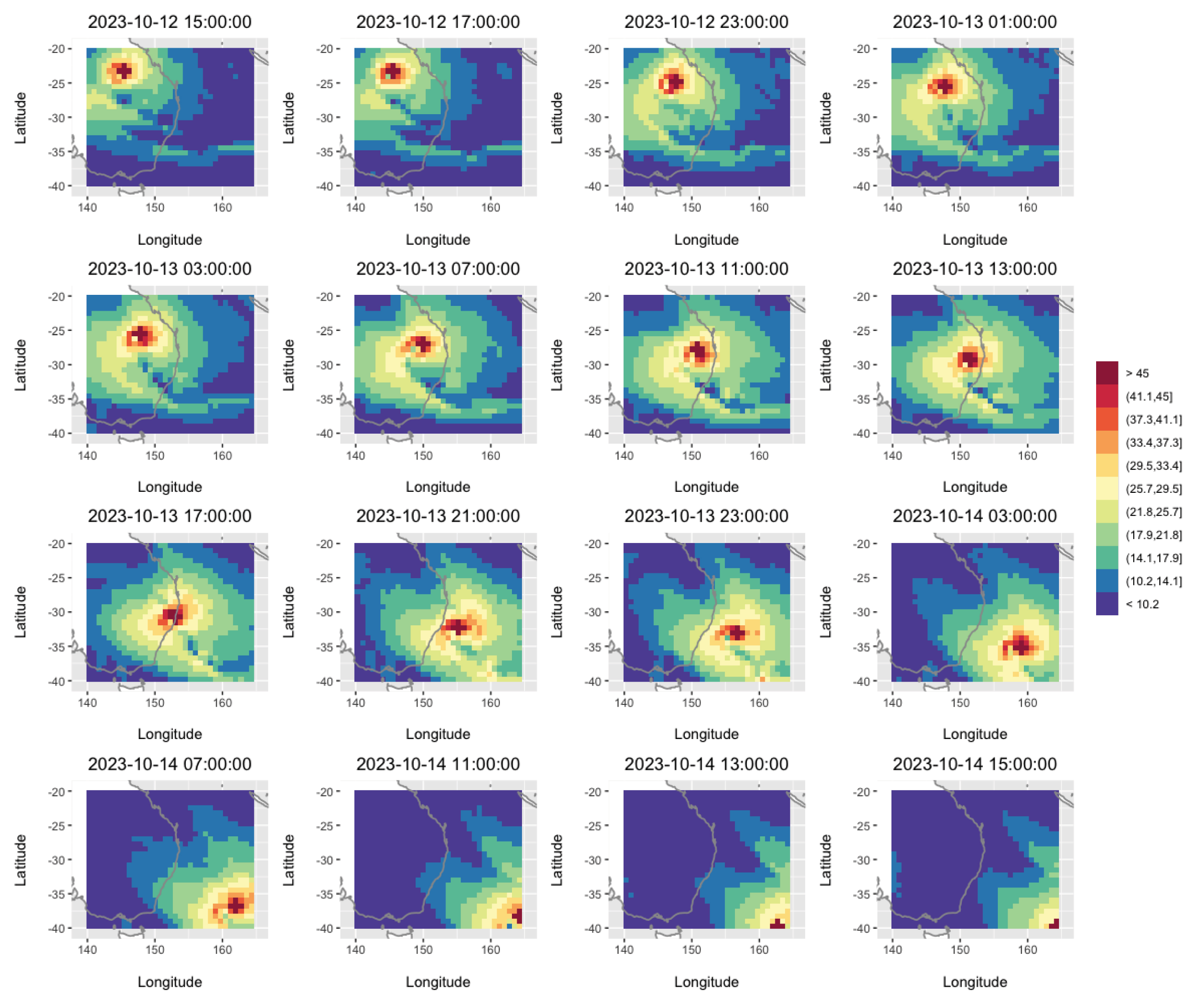}
  \caption{Snapshots of the simulated data over time. Extreme events are assumed at locations highlighted in dark red (i.e., $\text{values}>45$).}
    \label{simul_fig1}
\end{figure}
The second experiment aims to show that our model can detect extreme events even when the model is mis-specified. We use hourly wind speed data from the ERA5 database \citep{Hersbach2023}, ranging from 12 October 15:00 to 14 October 16:00, 2023 ($T=50$), near Australia. The data are upscaled to $N=891$ locations over $[140.25, 164.25]$ $\times$ $[-20.25, -39.75]$. To simulate extremes, we replace values near the maximum at each time point with values well beyond the typical range observed for ordinary events. Figure \ref{simul_fig1} shows spatial plots with extreme values (dark red) shifting from northwest to southeast. 

\begin{figure}[!t]
\centering
\includegraphics[width=1\linewidth]{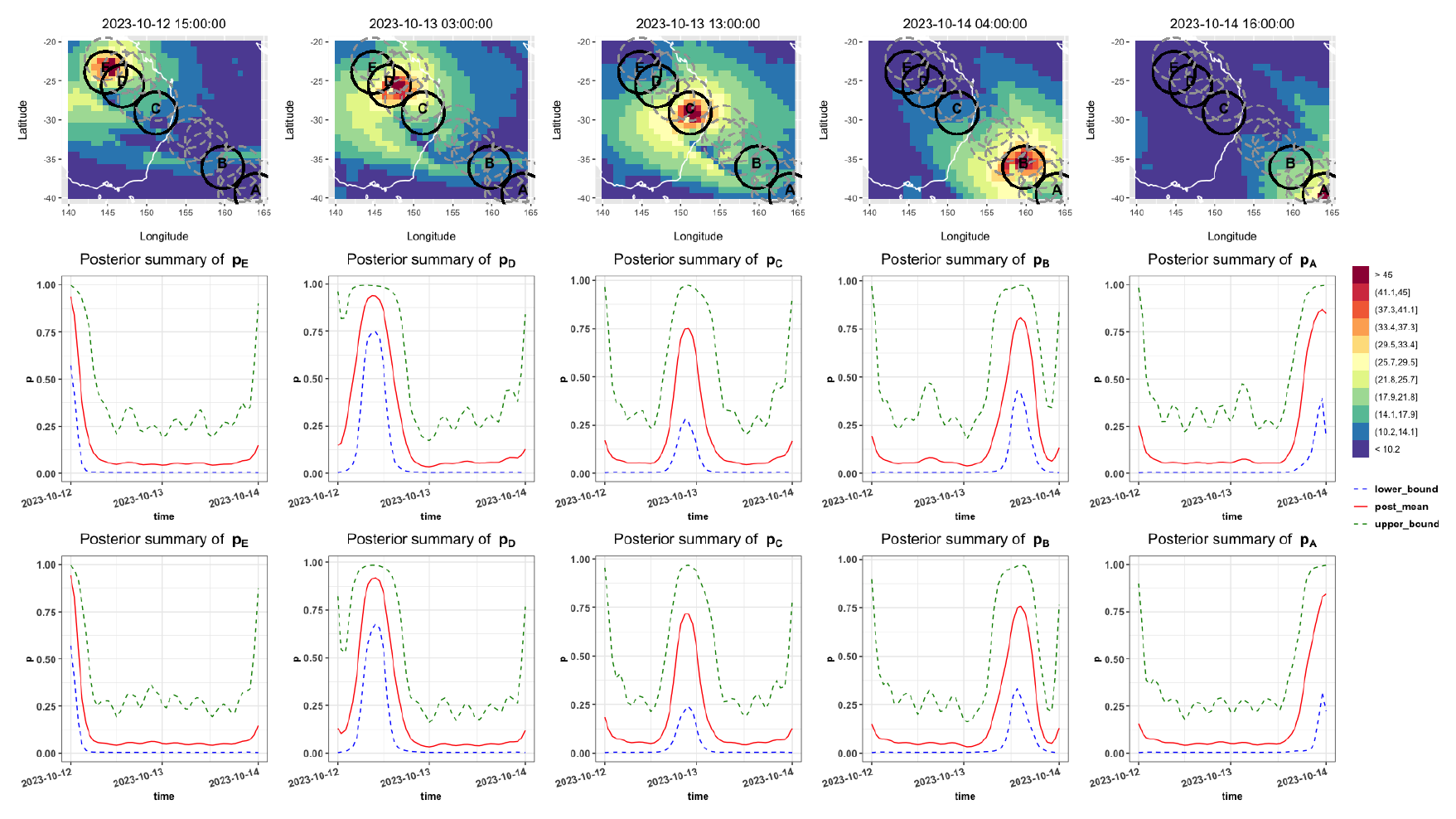} 
  \caption{Top: Regions covered by knots, indexed by characters, over the data. Regions where the posterior mean of $\bm{p}_t$ exceeds 0.7—based on the variance-gamma or stable distribution—are highlighted with either dashed or solid lines. Middle and Bottom: Posterior summaries of $\bm{p}_t$ at knot locations labeled ``A" through ``E" for the variance-gamma and stable distribution cases, respectively.}
    \label{simul_fig2}
\end{figure}

We fit two models using the variance-gamma and stable distributions, drawing 20,000 posterior samples and discarding the first 10,000 as burn-in. Matrices $\bm{\Phi}$ and $\bm{B}$ are constructed using Wendland and B-spline basis functions with $K = 144$ and $q = 10$ uniformly spaced knots, respectively. Priors for $\sigma_{\text{vg},k}$ and $\nu_{s,k}$ are $\text{Inv.gamma}(2, 100)$, while $\sigma^2_{\text{Gau},k} \sim \text{Inv.gamma}(2, 20)$. We set $\lambda_k \sim \text{TN}(1, 0.14, \ell=0, u=2)$ and $\theta_{\text{vg},k} \sim \text{Gau}(0, 25)$, with vague priors for remaining parameters.

Figure \ref{simul_fig2} presents posterior summaries of $p_{k,t}$ for selected $k$ and corresponding data at some $t$. Both models yield similar $\bm{p}_t$ distributions. Regions associated with knots $A$ through $D$ capture extremes at the correct times. For example, the posterior mean at ``B" exceeds 0.7 around 04:00 on 14 October 2023 but remains near zero otherwise. This demonstrates the model's ability to detect extremes and quantify uncertainty, even under model misspecification.

\section{$\text{PM}_{2.5}$ Data Analysis}\label{sec:6}
\begin{figure}
\centering
\includegraphics[width=1\linewidth]{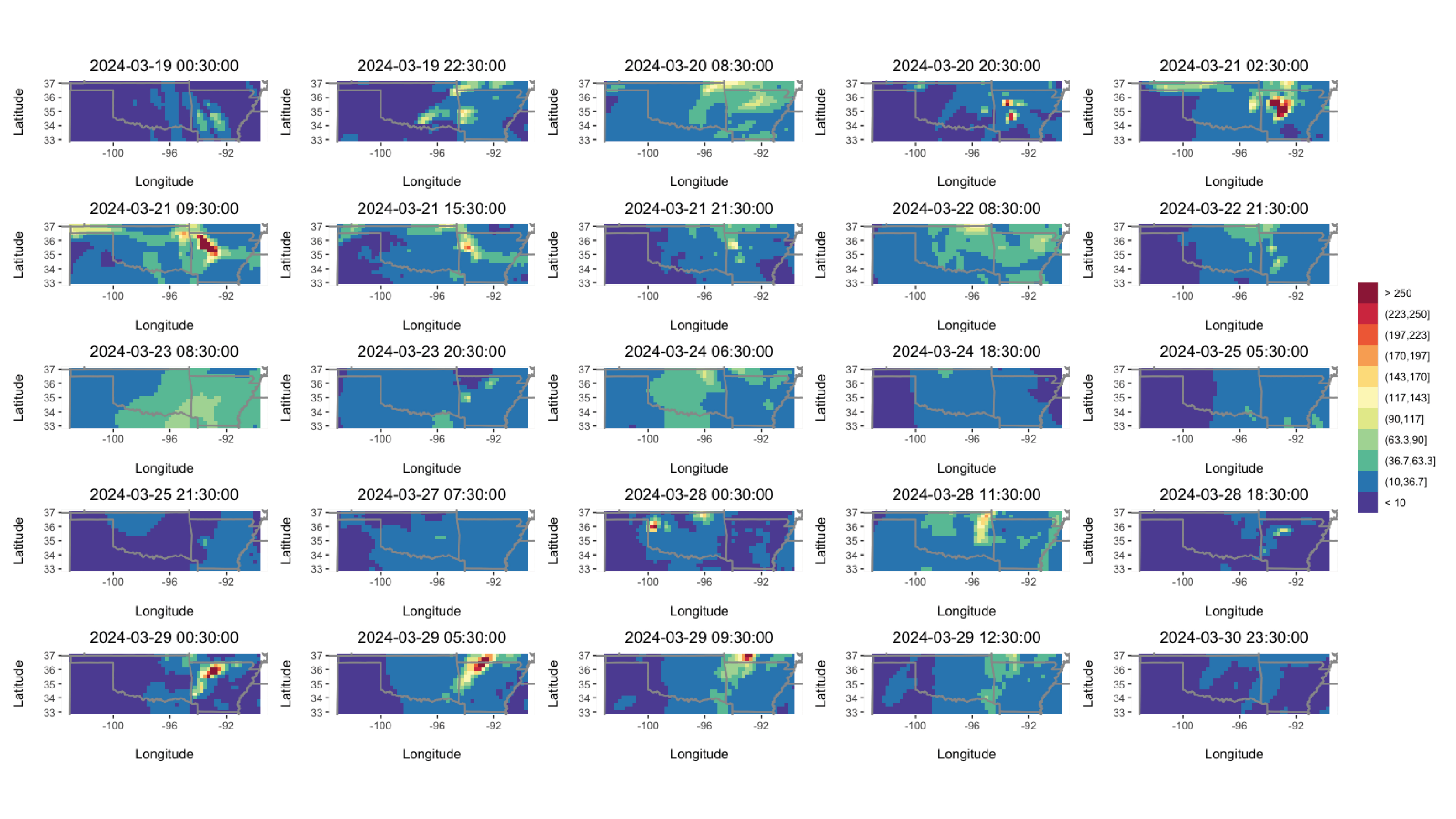}
  \caption{Snapshots of $\text{PM}_{2.5}$ across time. The concentration of PM2.5 increases significantly around 02:30 on March 21, 2024.}
    \label{fig4}
\end{figure}
To illustrate our model, we analyze hourly PM$_{2.5}$ concentrations over the Central U.S. in March 2024, using data from NASA’s GEOS-CF v1.0 system \citep{https://doi.org/10.1029/2020MS002413}\footnote{\url{https://gmao.gsfc.nasa.gov/weather_prediction/GEOS-CF/}}. Focusing on Arkansas, Oklahoma, and southwestern Missouri from March 19 at 00:30 to March 30 at 23:30, the dataset includes $N=918$ locations (25 km resolution) and $T=288$ time points (Figure \ref{fig4}). PM$_{2.5}$ are highly right-skewed, with a maximum of 756.09 \si{\micro\gram}$/m^3$ and 90\%, 95\%, and 99\% quantiles of 39.47, 50.65, and 90.10 \si{\micro\gram}$/m^3$. Extreme concentrations emerge near the Missouri–Arkansas border around 02:30 on March 21 and spread westward, likely driven by a fire in the region that day.\footnote{\url{https://modis.gsfc.nasa.gov/gallery/individual.php?db_date=2024-03-21}} For the analysis, the data are marginally transformed as $\widetilde{\bm{Y}}_t = \log(\exp(\bm{Y}_t) - 1)$ to match the model's support while preserving the original distribution's upper heavy tail.

\subsection{Detection of extreme events} \label{sec:4.6.2}
\begin{table}[!t]
\centering
\caption{Comparison between models using the stable or variance-gamma distribution in the regime-switching process. MSE, IS, and CRPS refer to the mean squared error, interval score, and continuous ranked probability score \citep{matheson1976scoring, Gneiting01032007}.}%
\label{table1}
\begin{tabular}{ccccccc}
\hline
Model & MSE & \textbf{} & IS &  & CRPS &  \\ \cline{1-2} \cline{4-4} \cline{6-6}
stable         & 68.195       &           & 37.653      &  & 2.261         &  \\
variance-gamma & 68.374       &           & 37.726      &  & 2.267         &  \\ \hline
\end{tabular}
\end{table}

To detect extreme events with uncertainty quantification, we fit the model using 170 uniformly spaced Wendland basis functions and $q=10$ B-spline temporal basis functions. We present results based on the stable distribution, which fits slightly better than the model with variance-gamma (Table \ref{table1}). Prior distributions are $\lambda_k \sim\text{TN}(\mu_{\lambda}=1,\sigma_{\lambda}=0.12,\ell=0,u=2)$, $\nu_{s,k} \sim \text{Inv.gamma}(\alpha_{\nu}=2.5,\beta_{\nu}=150)$, and $\sigma^2_{\text{Gau},k} \sim \text{Inv.gamma}(\alpha_g=2.5,\beta_g=30)$. For other parameters, we assign vague priors. With these priors and $\sigma^2_{\eta} = 0.5$, we draw 20,000 posterior samples, discarding the first 10,000 as burn-in.
\begin{figure}[!t]
\centering
\includegraphics[width=1\linewidth]{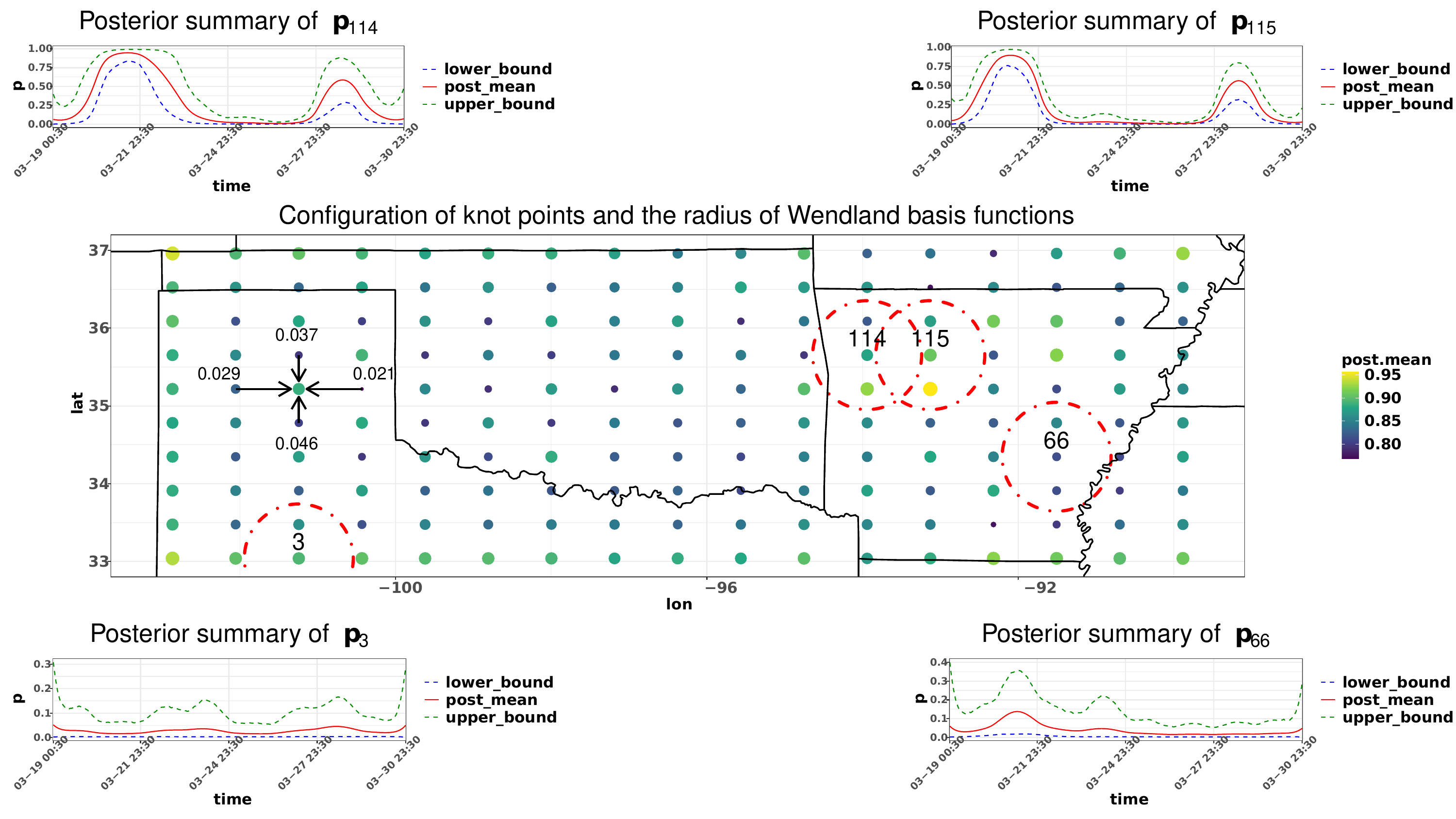}
  \caption{Illustration of the posterior PM$_{2.5}$ model. Top left, top right, bottom left, and bottom right displays: Posterior mean and 95\% credible intervals of tail-switching probabilities $p_{kt}$ across time $t$, for knot locations $k\in \{114,115,3,66\}$. Middle display: Knot positions for basis functions are indicated by dots whose sizes and colours indicate posterior means of the corresponding diagonal element of matrix $\bm M$, with the radius of basis functions represented by a dashed red line for the selected knots. Arrows (third column from the left)  show posterior means of the off-diagonal elements of $\bm M$ associated with each knot.}
    \label{fig7}
\end{figure}
\begin{figure}[!t]
\centering
\includegraphics[width=0.95\linewidth]{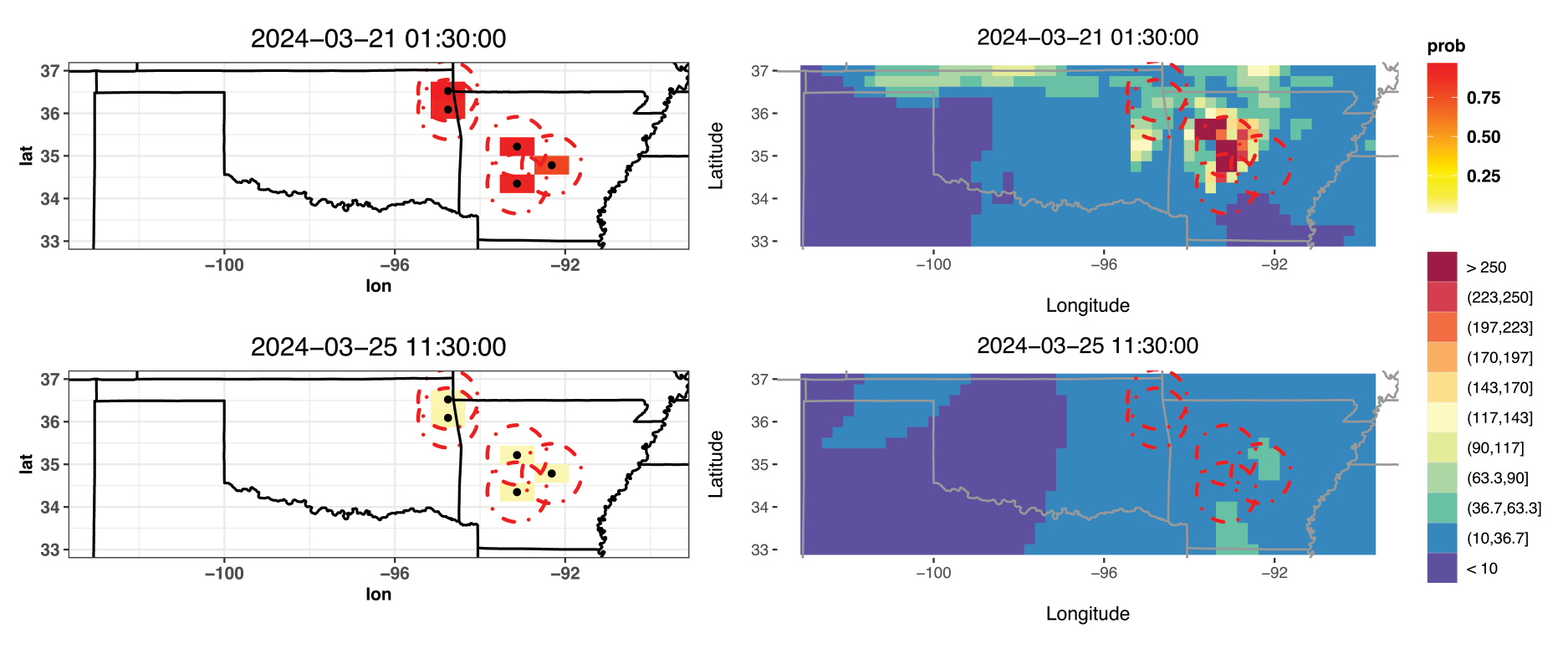}
  \caption{Left: the posterior mean of $p_{kt}$ at $t=$21 March, 2024, 01:30 and $t=$25 March, 2024, 11:30, for regions where the posterior mean exceeds 0.95 at some point during the study period. Right: True PM$_{2.5}$ at the corresponding times.}
    \label{fig8}
\end{figure}

Figure \ref{fig7} summarizes the posterior of $\hat{p}_{kt}$ over time and space for $k \in \{ 3,66,114,115\}$, with knot locations and radii from the Wendland basis. Extreme events are detected near $\bm{\Phi}_{\cdot,114}$ and $\bm{\Phi}_{\cdot,115}$, where posterior means exceed 0.89 around 00:30 on 20 March 2024. In contrast, $\bm{p}_3$ and $\bm{p}_{66}$ remain near zero, indicating no extreme events in those regions. As discussed in Section \ref{sec:example}, the transition matrix $\bm{M}$ quantifies the temporal propagation of extremes to neighbouring knots. As can be seen in Figure \ref{fig7}, the extreme at a knot at time $t+1$ is primarily driven by the same knot at $t$, while northward propagation is slightly more pronounced than in other directions.

Figure \ref{fig8} shows the posterior mean (left) and true PM${2.5}$ concentrations (right) at two time points, highlighting locations where $\hat{p}_{kt}$ exceeds 0.96 at some time point. High posterior means at 01:30 on March 21 and low values at 11:30 on March 25 align with observed PM$_{2.5}$ patterns, particularly in central and eastern Oklahoma and western Arkansas. The figures show the model’s ability to detect spatio-temporal extremes and represent associated uncertainty.

\subsection{Evaluation of predictive performance for times with missing data} 
\label{sec:4.6.1}

In the spatial and spatio-temporal extremes literature, prediction over the entire spatial domain at a missing time $t$ is rarely addressed, although it is often feasible using the models proposed therein. However, these models cannot fully recover the original process, as block maxima and peaks-over-threshold methods rely on only a subset of the data. In contrast, prediction at unobserved time points is straightforward in our model. To evaluate the model’s predictive performance, we exclude the data at 01:30 on March 21 and 03:30 on March 23, and fit the model with a stable distribution to the remaining data, using the same priors, basis functions, and posterior sample size as in Section \ref{sec:4.6.2}.
\begin{figure}[!t]
\centering
\includegraphics[width=1\linewidth]{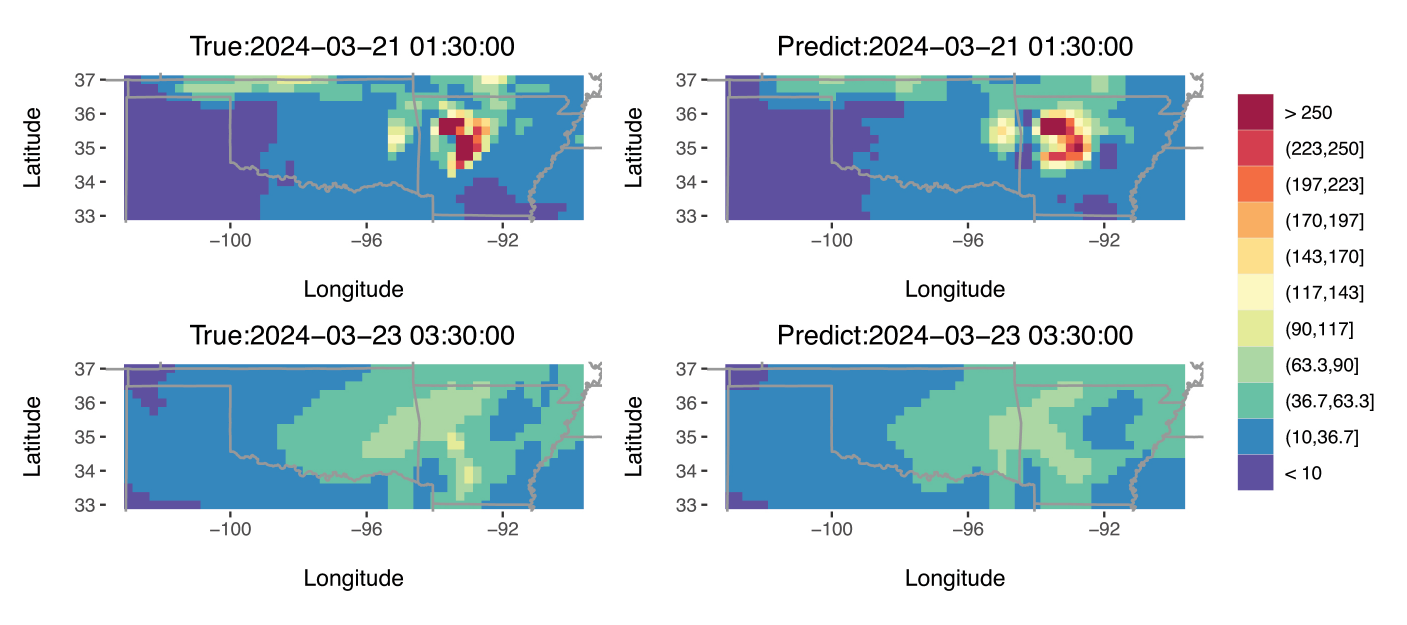}
  \caption{Top and bottom left: True PM${2.5}$ at 01:30 on March 21 and 03:30 on March 23.
    Top and bottom right: Posterior mean of the predicted PM${2.5}$ at the corresponding times.}
    \label{fig10}
\end{figure}

Figure \ref{fig10} presents the true $\text{PM}{2.5}$ concentrations (left) annd the posterior mean of the predictions (right). While the posterior mean with 170 basis functions is limited in capturing small scale variation in data, it can effectively represent larger-scale patterns and trends. At 01:30 on March 21, it identifies the high PM$_{2.5}$ region in central-west Arkansas but slightly underestimates its intensity. In contrast, at 03:30 on March 23, the posterior means of predictions align more closely with the true PM$_{2.5}$. This may be due to the inherent smoothing effect of the low-rank approximation based on basis functions. Increasing the number of basis functions may better capture small-scale variation, but at higher computational cost.

\section{Discussion and conclusion}
\label{sec:7}
Extreme events in environmental processes can lead to severe consequences, emphasizing the importance of understanding their dynamics and implications. These events are often identified using various science-motivated indices, typically without accounting for uncertainty. Although many statistical models, particularly in the spatial extremes literature, have been developed to better characterize complex environmental processes, many of them rely on restrictive assumptions and often fail to incorporate the underlying dynamic mechanisms into the model. In particular, the interaction of spatial and temporal dependence of extremes with flexible combinations of different dependence types remains largely unexplored. 

In this work, we consider a regime-switching process within a spatio-temporal dynamic model to address those limitations. Specifically, we employ a regime-switching process between Gaussian and heavier-tailed distributions to capture the effect of external forcing or innovations that may drive extreme events. This spatio-temporal dynamic model also readily captures spatial and temporal dependencies, as well as their interactions. The tail behaviour of the innovation variables, combined with the matrix $\bm M$, determines how asymptotic dependence of extremes propagates through space and time. In an application to hourly PM$_{2.5}$ concentration in the Central U.S. in March 2024, we demonstrate that our model can detect extreme events across space and time while providing associated uncertainty quantification. We also show that our model can be used to make predictions at a missing time $t$.

As discussed in Section~\ref{sec:example}, innovations with regularly varying tails produce persistent asymptotic dependence that propagates across both space and stretch over relatively periods before decaying to zero. By contrast, exponentially-tailed innovations yield AD locally in space and time. In practice, if the data margins are heavy-tailed but true AD is confined to small neighbourhoods, it is more suitable to instead adopt a ``multiplicative'' dynamic model of type $\exp(\bm Y_t)$ where $\bm Y_t$ is the dynamic model with exponentially-tailed innovations.

There are several avenues to extend this model. First, it would be desirable to incorporate covariates into the model. Along with the temporal basis functions and their coefficients, covariates can be used to approximate $p_{kt}$, for example by considering indices for large-scale atmospheric stability, circulation and teleconnection patterns. This would also facilitate inference on covariates, including quantifying their effects on the occurrence of extreme events. Second, accounting for potential non-linearity would make the model more realistic. Many environmental processes exhibit non-linearity, and accommodating this in the model would be an important step toward improving its capacity to capture complex system dynamics. Third, modeling extremes in the mean of the process rather than in the innovation may improve predictive performance. Although a regime-switching process in the innovations effectively detects extreme events, it does not necessarily improve predictive performance, as it affects innovations rather than the mean of the process model. Perhaps, it can be done by applying a regime-switching process on the mean term, as described similarly in \citet{shumway1991dynamic}. Moreover, marginal behaviour and dependence of tails are both determined by the innovation distribution in our model -- exponential tails and  tail dependence that is local in space-time, or regularly varying tails with possibly time-persisting tail dependence that, additionally, could be spatially nonlocal in case of nonlocal basis functions. To improve flexibility, one could allow for additional marginal transformations, for example using the multiplicative process $\exp(\bm Y_t)$ with exponential-tailed $\bm Y_t$ to accommodate heavy marginal tails with local-only tail dependence. 
These extensions would enhance the model’s flexibility and effectiveness for both detecting extreme events and making predictions. Incorporating covariates, nonlinearity, and improved representations of extreme behaviour would broaden the framework’s applicability to a wider range of environmental applications.

\newpage

\appendix
\renewcommand{\thesection}{\Alph{section}}

\section{Details regarding the example in Section~\ref{sec:example}}\label{sec:example_proof}
\subsection{Regularly varying innovations}\label{sec:example_RV}
For the example in Section~\ref{sec:example} under the RV innovations and $K=4$ local Wendland basis functions, we can write out the basis matrix explicitly:
\[
\bm\Phi=\left(
\begin{smallmatrix}
0 & 0 & 0 & 1 \\[1pt]
0 & 1 & 0 & 0 \\[1pt]
0 & 1/2 & 0 & 1/2 \\[1pt]
1/2 & 0 & 1/2 & 0 \\[1pt]
0 & 0 & 1 & 0
\end{smallmatrix}
\right).
\]

\noindent In addition, based on Figure~\ref{fig1_theory}, the transition matrix can be written as 
\[
\bm M=\left(
\begin{smallmatrix}
m_{11} & 0 & 0 & 0 \\
0 & m_{22} & m_{23} & 0 \\
0 & 0 & m_{33} & 0 \\
0 & 0 & 0 & m_{44} \\
\end{smallmatrix}
\right)
\]
where all the non-zero elements are in $(0,1)$. It is easy to derive
\[
\bm M^2=\left(
\begin{smallmatrix}
m_{11}^2 & 0 & 0 & 0 \\
0 & m_{22}^2 & m_{22}m_{23}+m_{23}m_{33} & 0 \\
0 & 0 & m_{33}^2 & 0 \\
0 & 0 & 0 & m_{44}^2 \\
\end{smallmatrix}
\right),
\]
and by induction, we know the diagonal elements of $\bm M^t$ for any $t\in \mathbb{N}^+$ consist of $\{m_{kk}^t: k=1,\ldots, 4\}$. Then we specifically write out $\bm\Psi_2 = \bm\Phi(\bm M^{2}, \bm M,\bm I)$:
\begin{equation}
    \bm\Psi_2 = 
    \left(
\begin{smallmatrix}
0 & 0 & 0 & m_{44}^2 & 0 & 0 & 0 & m_{44} & 0 & 0 & 0 &1\\
0 & m_{22}^2&m_{22}m_{23}+m_{23}m_{33} & 0 & 0 & m_{22} & m_{23} & 0 & 0 & 1 & 0 & 0\\
0 & \frac{m_{22}^2}{2}&\frac{m_{22}m_{23}+m_{23}m_{33}}{2} & \frac{m_{44}^2}{2} & 0 & \frac{m_{22}}{2} & \frac{m_{23}}{2} & \frac{m_{44}}{2} & 0 & \frac{1}{2} & 0 & \frac{1}{2}\\
\frac{m_{11}^2}{2} & 0 &\frac{m_{33}^2}{2} & 0 & \frac{m_{11}}{2} & 0 & \frac{m_{33}}{2} & 0 & \frac{1}{2} & 0 & \frac{1}{2} & 0\\
0 & 0 & m_{33}^2 & 0 & 0 & 0 & m_{33} & 0 & 0 & 0 & 1 & 0\\
\end{smallmatrix}
\right).
\end{equation}

Since there is only one heavy-tailed RV innovation imposed on the $1^{\rm st}$--$3^{\rm rd}$ local basis function at time $0$, we simply need to focus on the first three columns of $\bm\Psi_t$ when examining the extremal dependence of $\chi(Y_{\bm s_i,t},Y_{\bm s_j,t})$ for $i\neq j$ and $t=0,1,2$. For instance, for $\bs_4$ and $\bs_5$, we know the $4^{\rm th}$ and $5^{\rm th}$ rows of $\bm\Psi_t$ start with $(\bm\Psi_t)_{4, \cdot}=(m_{11}^t/2, 0, m_{33}^t/2, \ldots)$ and $(\bm\Psi_t)_{5, \cdot}=(0, 0, m_{33}^t, \ldots)$ for $t\in \mathbb{N}^+$. By Theorems~\ref{theorem1} and \ref{theorem2}, it is easy to see $J_A={3}$ in these cases and 
\begin{equation*}
    \chi(Y_{\bm s_4,t},Y_{\bm s_5,t})=\min\left\{\frac{(m_{33}^t\nu_3)^{\lambda}(1+\kappa_3)}{(m_{11}^t\nu_1)^{\lambda}(1+\kappa_1)+(m_{33}^t\nu_3)^{\lambda}(1+\kappa_3)}, \frac{(m_{33}^t\nu_3)^{\lambda}(1+\kappa_3)}{(m_{33}^t\nu_3)^{\lambda}(1+\kappa_3)}\right\}, 
\end{equation*}
where the second term on the right-hand side is $1$, and hence \eqref{eqn:RV_example} holds.

To evaluate cross–temporal extremal dependence, we again trace the
coefficients that multiply the initial shocks
$\omega_{10},\,\omega_{20},\,\omega_{30}$ as they propagate to the times of
interest.  If the index set of non‑zero coefficients, $J_A$, is
non‑empty, the two variables share at least one common driver and
therefore exhibit positive same‑tail dependence.  For example, the pair
$\bigl(Y_{\bm s_2,1},\,Y_{\bm s_3,2}\bigr)$ inherits strictly
positive weights from $\omega{20}$ and $\omega{30}$, giving
$J_A=\{2,3\}$.  All of these weights are non‑negative, so
$\chi^{UU}>0$ and $\chi^{LL}>0$, whereas the opposite‑tail
coefficients vanish $\chi^{LU}=\chi^{UL}=0$.

\subsection{Exponentially-tailed innovations}\label{sec:example_ET}
\begin{figure}[H]
    \centering
    \includegraphics[width=0.25\linewidth]{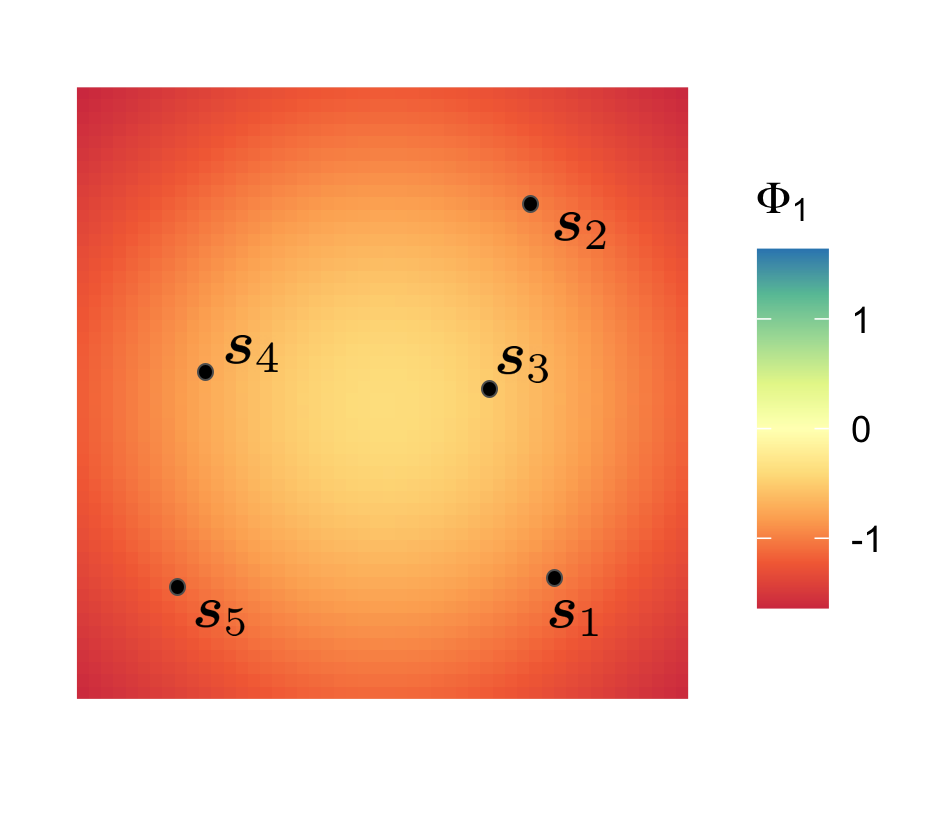}
     \hspace*{-0.3cm}\includegraphics[width=0.25\linewidth]{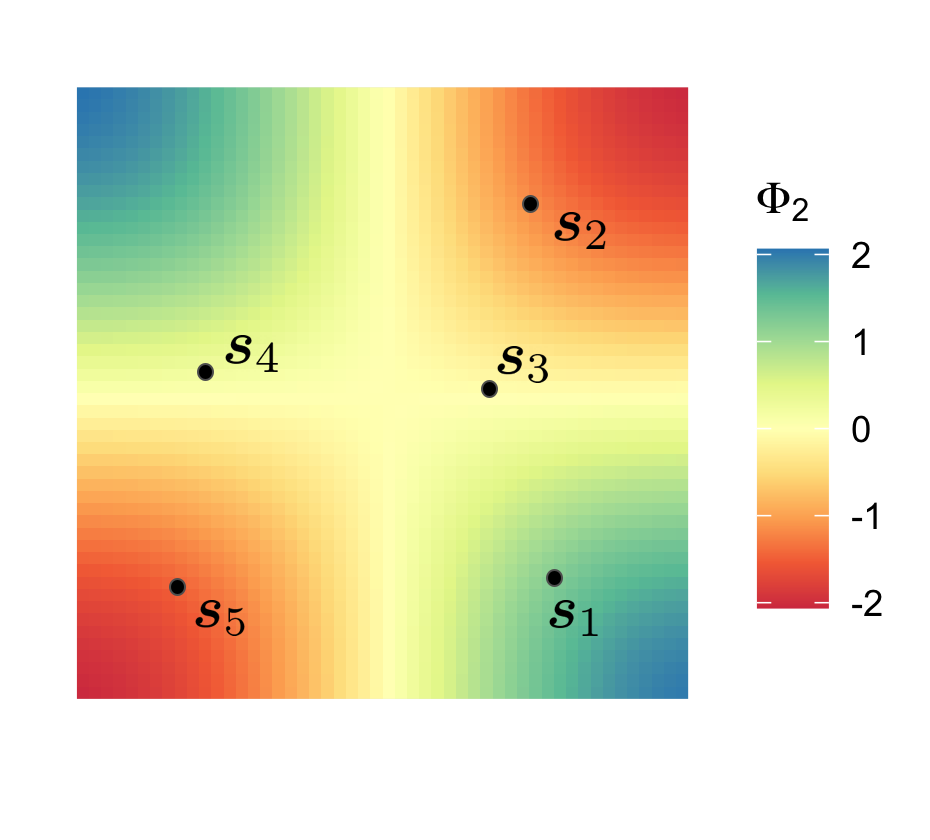}
      \hspace*{-0.3cm}\includegraphics[width=0.25\linewidth]{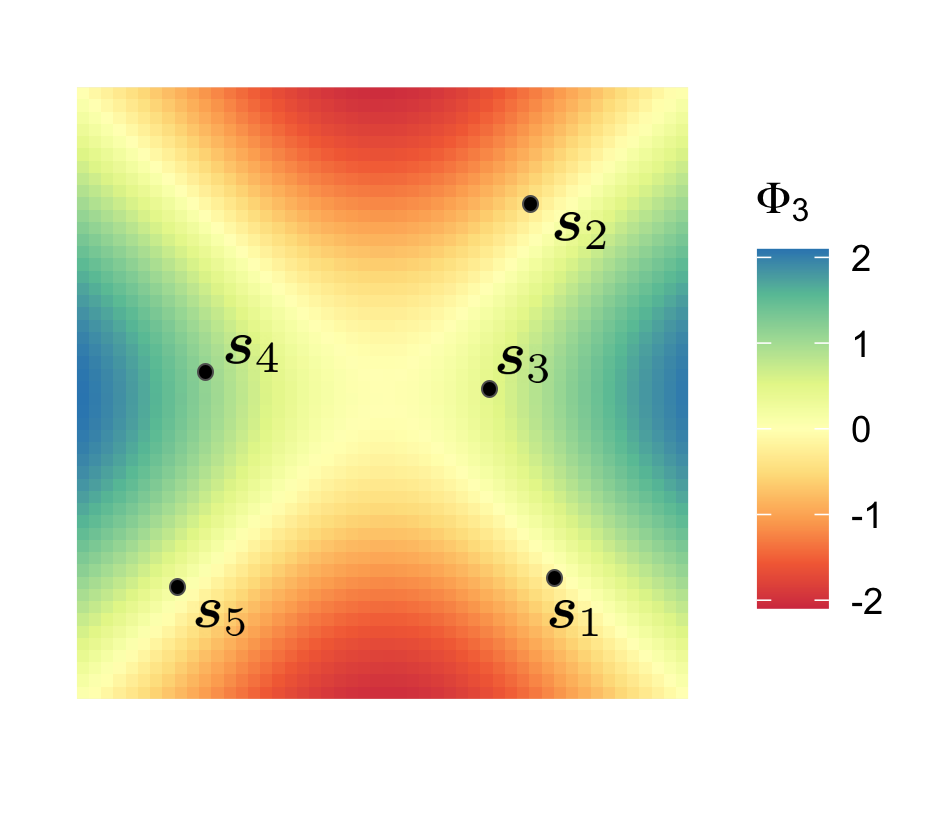}
       \hspace*{-0.3cm}\includegraphics[width=0.25\linewidth]{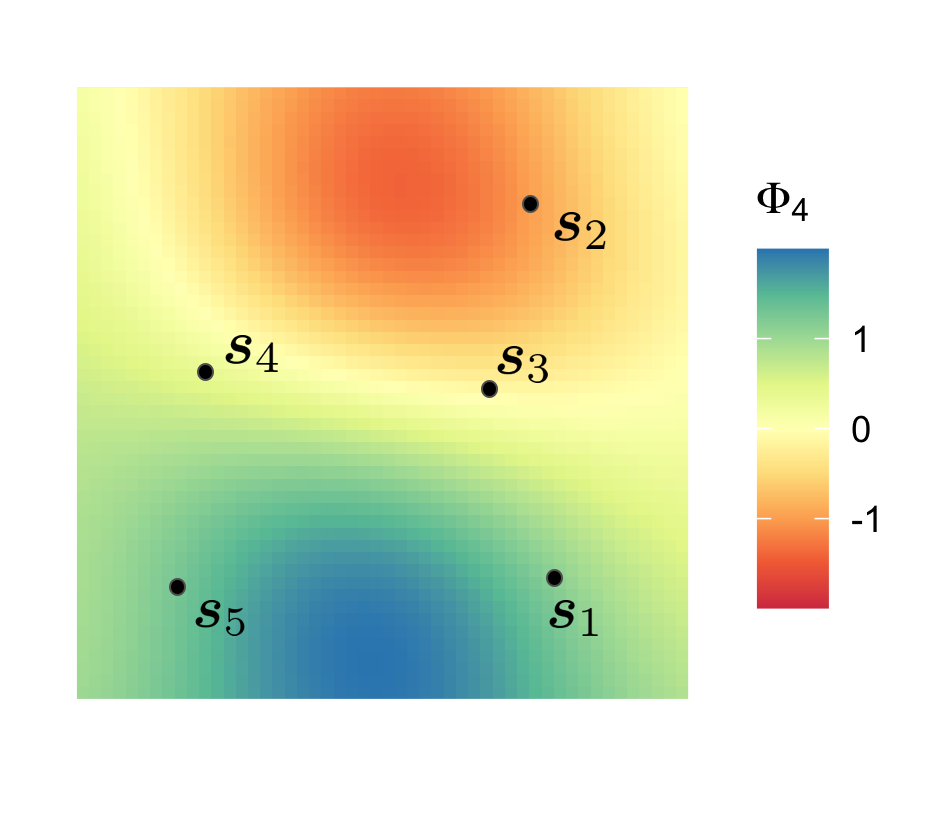}
    \caption{The graphical illustration of Theorem~\ref{theorem1} with $J=4$ global thin‐plate spline basis functions $\bm\Phi_{\cdot,1},\ldots,\bm\Phi_{\cdot,4}$, depicted in the four displays above. The locations are placed in the same way as  Figure~\ref{fig1_theory} in which we use local basis functions.}
    \label{fig:global_basis}
\end{figure}

In Figure~\ref{fig:global_basis}, we show the first four thin-plate spline basis functions on the square domain generated with the \texttt{R} package \texttt{mgcv} \citep{wood2015package}. For the five marked locations, the basis matrix can be written as
\[
\bm\Phi=\left(
\begin{smallmatrix}
\negg{\phi_{11}} & \pos{\phi_{12}} & \negg{\phi_{13}} & \pos{\phi_{14}} \\
\negg{\phi_{21}} & \negg{\phi_{22}} & \negg{\phi_{23}} & \negg{\phi_{24}} \\
\negg{\phi_{31}} & \negg{\phi_{32}} & \pos{\phi_{33}} & \negg{\phi_{34}} \\
\negg{\phi_{41}} & \pos{\phi_{42}} & \pos{\phi_{43}} & \pos{\phi_{44}} \\
\negg{\phi_{51}} & \negg{\phi_{52}} & \pos{\phi_{53}} & \pos{\phi_{54}}
\end{smallmatrix}
\right),
\]
in which the colors of the basis elements are consistent with the signs of the basis elements shown in Figure~\ref{fig:global_basis}. For this case, we assume the transition matrix to be
\[
\bm M=\left(
\begin{smallmatrix}
0 & 0 & 0 & 0 \\
m_{21} & 0 & 0 & 0 \\
0 & m_{32} & 0 & 0 \\
0 & 0 & m_{43} & 0 \\
\end{smallmatrix}
\right).
\]
Therefore, $\bm M$ propagates the process from time $t$ to $t+1$ via  $\bm\Phi_{\cdot,1}\rightarrow\bm\Phi_{\cdot,2}\rightarrow\bm\Phi_{\cdot,3}\rightarrow\bm\Phi_{\cdot,4}$; that is, the dynamic system only passes along extremes from larger scales (lower spatial frequency basis functions) to smaller spatial scales. This is analogous to the well-known forward energy cascade in turbulent flows, where large scales pass energy to smaller scales (e.g., large eddies break down to smaller scales and eventually dissipate). Also, it is easy to show
\[
\bm M^2=\left(
\begin{smallmatrix}
0 & 0 & 0 & 0 \\
0 & 0 & 0 & 0 \\
m_{32}m_{21} & 0 & 0 & 0 \\
0 & m_{43}m_{32} & 0 & 0 \\
\end{smallmatrix}
\right),\;
\bm M^3=\left(
\begin{smallmatrix}
0 & 0 & 0 & 0 \\
0 & 0 & 0 & 0 \\
0 & 0 & 0 & 0 \\
m_{43}m_{32}m_{21} & 0 & 0 & 0 \\
\end{smallmatrix}
\right),
\]
and $\bm M^t=\bm 0$ for any $t\geq 4$. 

In addition, we assume that the exponentially-tailed innovation only happens at the third basis frequency at time 0, i.e., $\omega_{30}\sim \text{VG}(\xi,\theta,\sigma,\mu)$ with $\lambda^{+}
=
\frac{\sqrt{\theta^2 + \sigma^2} - \theta}{\sigma^2}$ and $\lambda^{-}
=
\frac{\sqrt{\theta^2 + \sigma^2} + \theta}{\sigma^2}$, and all the other innovations are normally distributed, i.e., their $\lambda^{+}$ and $\lambda^{-}$ are $\infty$. To derive the $\chi$-coefficients for $t=0,1,2$, we specifically write out $\bm\Psi_2 = \bm\Phi(\bm M^{2}, \bm M,\bm I)$:
\begin{equation}
    \bm\Psi_2 = 
    \left(
\begin{smallmatrix}
m_{32}m_{21}\negg{\phi_{13}} & m_{43}m_{32}\pos{\phi_{14}} & 0 & 0 & m_{21}\pos{\phi_{12}} & m_{32}\negg{\phi_{13}} & m_{43}\pos{\phi_{14}} & 0 & \negg{\phi_{11}} & \pos{\phi_{12}} & \negg{\phi_{13}} & \pos{\phi_{14}} \\
m_{32}m_{21}\negg{\phi_{23}} & m_{43}m_{32}\negg{\phi_{24}} & 0 & 0 & m_{21}\negg{\phi_{22}} & m_{32}\negg{\phi_{23}} & m_{43}\negg{\phi_{24}} & 0 & \negg{\phi_{21}} & \negg{\phi_{22}} & \negg{\phi_{23}} & \negg{\phi_{24}} \\
m_{32}m_{21}\pos{\phi_{33}} & m_{43}m_{32}\negg{\phi_{34}} & 0 & 0 & m_{21}\negg{\phi_{32}} & m_{32}\pos{\phi_{33}} & m_{43}\negg{\phi_{34}} & 0 & \negg{\phi_{31}} & \negg{\phi_{32}} & \pos{\phi_{33}} & \negg{\phi_{34}} \\
m_{32}m_{21}\pos{\phi_{43}} & m_{43}m_{32}\pos{\phi_{44}} & 0 & 0 & m_{21}\pos{\phi_{42}} & m_{32}\pos{\phi_{43}} & m_{43}\pos{\phi_{44}} & 0 & \negg{\phi_{41}} & \pos{\phi_{42}} & \pos{\phi_{43}} & \pos{\phi_{44}} \\
m_{32}m_{21}\pos{\phi_{53}} & m_{43}m_{32}\pos{\phi_{54}} & 0 & 0 & m_{21}\negg{\phi_{52}} & m_{32}\pos{\phi_{53}} & m_{43}\pos{\phi_{54}} & 0 & \negg{\phi_{51}} & \negg{\phi_{52}} & \pos{\phi_{53}} & \pos{\phi_{54}}
\end{smallmatrix}
\right),
\end{equation}
We denote the sequence of integers from $n$ to a larger integer $N$ by $[\![n,N]\!]$, and denote the set without the element $n<m<N$ by $[\![n,N]\!]\setminus m$. Then by Theorem~\ref{thm:ET}, we can derive the $\chi$-coefficient for any space-time pairs. However, for the brevity of this Supplementary Material, we focus on $\bm s_1$ and $\bm s_2$ at different times. The other cases can be derived following the same procedure outlined in Theorem~\ref{thm:ET}. Firstly, at time $0$,
\begin{equation*}
    \begin{split}
        \chi^{UU}(Y_{\bm s_1,0},Y_{\bm s_2,0}) &= \mathbb{E}\left\{\min_{i\in\{1,2\}}\frac{\exp\left(\sum_{j\in [\![1,4]\!]\setminus 3}\lambda^-\phi_{ij}\omega_{j0}/|\phi_{i3}|\right)}{\mathbb{E}\exp\left(\sum_{j\in [\![1,4]\!]\setminus 3}\lambda^-\phi_{ij}\omega_{j0}/|\phi_{i3}|\right)}\right\},\\
        \chi^{LL}(Y_{\bm s_1,0},Y_{\bm s_2,0}) &= \mathbb{E}\left\{\min_{i\in\{1,2\}}\frac{\exp\left(\sum_{j\in [\![1,4]\!]\setminus 3}\lambda^+\phi_{ij}\omega_{j0}/|\phi_{i3}|\right)}{\mathbb{E}\exp\left(-\sum_{j\in [\![1,4]\!]\setminus 3}\lambda^+\phi_{ij}\omega_{j0}/\phi_{i3}\right)}\right\},
    \end{split}
\end{equation*}
and $\chi^{LU}(Y_{\bm s_1,0},Y_{\bm s_2,0})=\chi^{UL}(Y_{\bm s_1,0},Y_{\bm s_2,0})=0$ because $\phi_{13}$ and $\phi_{23}$ are both negative.

Interestingly, as we progress from time $0$ to time $1$, the coefficients multiplied onto $\omega_{30}$ become the $7^{\rm th}$ column of $\bm\Psi_2$ (i.e., the $3^{\rm rd}$ column of $\bm\Psi_1$), and we see that $m_{43}\phi_{14}$ and $m_{43}\phi_{24}$ now have opposite signs. Therefore, we have
\begin{footnotesize}
    \begin{equation*}
    \begin{split}
        \chi^{LU}(Y_{\bm s_1,1},Y_{\bm s_2,1}) &= \mathbb{E}\left[\min_{i\in\{1,2\}}\frac{\exp\{\lambda^-\left(\sum_{j\in [\![1,2]\!]}m_{j+1,1}\phi_{i,j+1}\omega_{j0}+\sum_{j\in [\![1,4]\!]}\phi_{ij}\omega_{j1}\right)/(m_{43}|\phi_{i4}|)\}}{\mathbb{E}\exp\{\lambda^-\left(\sum_{j\in [\![1,2]\!]}m_{j+1,1}\phi_{i,j+1}\omega_{j0}+\sum_{j\in [\![1,4]\!]}\phi_{ij}\omega_{j1}\right)/(m_{43}|\phi_{i4}|)\}}\right],\\
        \chi^{UL}(Y_{\bm s_1,1},Y_{\bm s_2,1}) &= \mathbb{E}\left[\min_{i\in\{1,2\}}\frac{\exp\{\lambda^+\left(\sum_{j\in [\![1,2]\!]}m_{j+1,1}\phi_{i,j+1}\omega_{j0}+\sum_{j\in [\![1,4]\!]}\phi_{ij}\omega_{j1}\right)/(m_{43}|\phi_{i4}|)\}}{\mathbb{E}\exp\{\lambda^+\left(\sum_{j\in [\![1,2]\!]}m_{j+1,1}\phi_{i,j+1}\omega_{j0}+\sum_{j\in [\![1,4]\!]}\phi_{ij}\omega_{j1}\right)/(m_{43}|\phi_{i4}|)\}}\right],
    \end{split}
\end{equation*}
\end{footnotesize}
whereas $\chi^{UU}(Y_{\bm s_1,0},Y_{\bm s_2,0})=\chi^{LL}(Y_{\bm s_1,0},Y_{\bm s_2,0})=0$.

At time 2, the third column of $\bm \Psi_2$ is $\bm 0$, and the effect of $\omega{30}$ is not propagated to this time, and therefore $\chi^{UU}(Y_{\bm s_1,2},Y_{\bm s_2,2})=\chi^{LL}(Y_{\bm s_1,2},Y_{\bm s_2,2})=\chi^{LU}(Y_{\bm s_1,2},Y_{\bm s_2,2})=\chi^{UL}(Y_{\bm s_1,2},Y_{\bm s_2,2})=0$.

Similarly, to examine the cross-time dependence, we just need to find the corresponding coefficients for $\omega{30}$ at different times and locations. For example, for $Y_{\bm s_1,0}$ and $Y_{\bm s_2,1}$, the desired coefficients are $\phi_{13}$ and $m_{43}\phi_{24}$, respectively. Since they have the same sign, we know $\chi^{UU}>0$ and $\chi^{LL}>0$ while $\chi^{LU}=\chi^{UL}=0$.

\begin{table}[]
\caption{The $\chi$-dependence coefficients for the configuration depicted in Figure~\ref{fig:global_basis}, where extreme innovations (exponentially-tailed) occur only at knots $3$ at the initial time point $t=0$. We report whether the coefficients are zero (i.e., AI) or positive (i.e., AD) in this setting. See Section \ref{sec:example_ET} of the Supplementary Material for the specific $\chi$-values for the AD cases.}
\label{tab:example_ET}
\centering
\begin{tabular}{c|cccc}
\hline
                            & $\chi^{UU}$ & $\chi^{LL}$ & $\chi^{LU}$ & $\chi^{UL}$ \\ \hline
$Y(\bm s_1,0),Y(\bm s_2,0)$ & AD          & AD          & AI          & AI          \\ \hline
$Y(\bm s_1,1),Y(\bm s_2,1)$ & AI          & AI          & AD          & AD          \\ \hline
$Y(\bm s_1,2),Y(\bm s_2,2)$ & AI          & AI          & AI          & AI          \\ \hline
$Y(\bm s_1,0),Y(\bm s_2,1)$ & AD          & AD          & AI          & AI          \\ \hline
$Y(\bm s_1,0),Y(\bm s_2,2)$ & AI          & AI          & AI          & AI          \\ \hline
\end{tabular}
\end{table}
In Table~\ref{tab:example_ET}, we summarize the dependence class for different space-time pairs. We note that the opposite-tail dependence does not co-exist with the same-tail dependence because we only have a single exponentially-tailed innovation at one time. One can easily come up with a more complex setting with multiple exponentially-tailed innovations at different times to exhibit a more flexible dependence structure.

\section{Linear combinations of regularly varying innovations}\label{proof:RV}

In this section, we examine the extremal dependence of a general bivariate random vector $(X_1,X_2)^{\top}$ formulated in~\eqref{eqn:additive_struc}, which may represent any space–time pairs. Assuming the innovation terms are regularly varying, we can derive extremal dependence properties, such as the coefficient $\chi$, from the spectral measure with the help of results such as Corollary~1 of \citet{Cooley2019}.  Due to the heavy-tail property, the distribution in the bulk of the variables $w_j$ will not influence joint tail properties in the dynamic model, which are fully determined by the univariate tails of $w_j$.

\begin{lemma}[\citeauthor{Cooley2019}, \citeyear{Cooley2019}]\label{lem:cooley-thibaud} 
Assume $\bm\omega=(w_1,\ldots, w_J)^{\top}$ are independently and identically distributed random variables that are regularly varying at infinity with the same index $\lambda>0$; equivalently, there exists a sequence $b_n\rightarrow\infty$ such that 
\begin{equation}\label{eq:rv-marg}
n \Pr(w_j/b_n > x) \rightarrow x^{-\lambda}, \quad n\rightarrow\infty,
\end{equation}
for any $j=1,\ldots, J$. Also, $\bm\Psi$ is a matrix with $m$ rows, $J$ columns and has only non-negative entries and at least one positive entry in each column. 

Then the random vector $\bm\Psi \bm \omega$ is also regularly varying at infinity with tail index $\lambda$ and it has angular measure (in dimension $m$ and with respect to any chosen norm $\|\cdot\|$) given by
\begin{equation}\label{eq:angular-measure-general}
H_{\bm\Psi \bm\omega}(\cdot) = \sum_{j=1}^J\left\|\bm\Psi_{\cdot,j}\right\|^\lambda \times \delta_{\bm\Psi_{\cdot,j}/\|\bm\Psi_{\cdot,j}\|}(\cdot),
\end{equation}
where $\bm\Psi_{\cdot,j}$ is the $j^{\rm th}$ column of $\bm\Psi$, $\delta(\cdot)$ is the Dirac mass function, and we set the sum term for $j$  to $0$ if $\left\|\bm\Psi_{\cdot,j}\right\|=0$.
\end{lemma}
Although this result is powerful, its assumptions are too restrictive for our setting. It presumes identically distributed RV components and permits only non‑negative weights in $\bm \Psi$, thereby describing upper‑tail behaviour alone. We now slightly reformulate the result of \citet{Cooley2019} where it accommodates generally independent (not necessarily identical) stable variables and admits weights of either sign, so that both upper‑ and lower‑tail dependence can be analysed. 

\begin{proposition}
Let $\bm \omega$ be a random vector with independently distributed components $\omega_{j}\sim \mathcal{S}(\lambda,\kappa_j,\nu_j,\delta_j)$ for $j=1,\ldots,J$ with $\kappa_j\in(-1,1)$ and $\lambda\in(0,2)$. Given a non null matrix $\bm\Psi$ (whose elements are not necessarily non-negative) and $\bm x = \bm \Psi\bm\omega$, the angular measure of $\bm x$ is
\begin{equation}\label{eqn:full_angular_measure}
    H_{\bm x}(\cdot) = \sum_{j=1}^J \left\|{\bm\psi}^+_j\right\|^\lambda \times \delta_{{\bm\psi}^+_j/\|{\bm\psi}^+_j\|}(\cdot) +  \sum_{j=1}^J \left\|{\bm\psi}^-_j\right\|^\lambda \times \delta_{{\bm\psi}^-_j/\|{\bm\psi}^-_j\|}(\cdot),
\end{equation}
where ${\bm\psi}^+_j=\max(\bm 0,\nu_j(1+\kappa_j)^{1/\lambda}\bm\Psi_{\cdot,j})$ and ${\bm\psi}^-_j=-\min(\bm 0,\nu_j(1-\kappa_j)^{1/\lambda}\bm\Psi_{\cdot,j})$, with the $\min$- and $\max$-operators applied componentwise, and where a contribution to the sum is defined as $0$ when ${\bm\psi}^{\pm}_j=\bm 0$.
\end{proposition}

\begin{proof}
    First, note that the upper and lower tails of the stable distribution are equivalent up to a positive scaling constant. Although the results in \citet{Cooley2019} only hold for a regularly varying random vector that takes values in $[0,\infty)^J$, the joint upper tail behaviour could be determined by both positive weights $\psi_{ij}$ multiplied with values in the upper tail of the components of ${\bm \omega}$ and negative weights $\psi_{ij}$ multiplied with the lower tail of the components of ${\bm \omega}$.  

Specifically, if we denote $\tilde{\omega}_j=\omega_j/\{\nu_j(1+\kappa_j)^{1/\lambda}\}$, we have by Lemma~\ref{lem:mar_survival} that $\Pr(\tilde{\omega}_j>x)\sim c_\lambda x^{-\lambda}$, $j=1,\ldots, J$. That is, $\tilde{\omega}_j$'s have identical tail behaviours. We also define new weights $\tilde{\psi}_{ij}=\psi_{ij}\nu_j(1+\kappa_j)^{1/\lambda}$, $i\in\{1,\ldots, m\}$ and $j\in \{1,\ldots,J\}$. Then $\bm \Psi\bm\omega=\tilde{\bm \Psi}\tilde{\bm\omega}$. We can therefore choose 
$$
b_n=n^{1/\lambda}c_\lambda^{1/\lambda}
$$
such that $n \Pr(\tilde{\omega}_j/b_n > x) \rightarrow x^{-\lambda}$, $n\rightarrow\infty$ and Lemma~\ref{lem:cooley-thibaud} holds for $\tilde{\bm\omega}$. Given a non-negative matrix $\bm\Psi$ and $\bm x = \bm \Psi\bm\omega$, the angular measure of $\bm x=\tilde{\bm \Psi}\tilde{\bm\omega}$ is
$$
H_{\bm x}(\cdot) = \sum_{j=1}^J \left\|\tilde{\bm\psi}_j\right\|^\lambda \times \delta_{\tilde{\bm\psi}_j/\|\tilde{\bm\psi}_j\|}(\cdot),
$$
where a contribution to the sum is defined as $0$ when $\tilde{\bm\psi}_j=\bm 0$.

Similarly, for the lower tail, we define $\tilde{\omega}^-_j=-\omega_j/\{\nu_j(1-\kappa_j)^{1/\lambda}\}$ and $\Pr(\tilde{\omega}^-_j>x)\sim c_\lambda x^{-\lambda}$, $j=1,\ldots, J$. Then we define new weights $\tilde{\psi}^-_{ij}=\psi_{ij}\nu_j(1-\kappa_j)^{1/\lambda}$, $i\in\{1,2\}$ and $j\in \{1,\ldots,J\}$. Then $\bm \Psi\bm\omega=-\tilde{\bm \Psi}^-\tilde{\bm\omega}^-$. Under the same $b_n$ as the previous case,
$$
H_{\bm x}(\cdot) = \sum_{j=1}^J \left\|\tilde{\bm\psi}^-_j\right\|^\lambda \times \delta_{\tilde{\bm\psi}^-_j/\|\tilde{\bm\psi}^-_j\|}(\cdot),
$$
where a contribution to the sum is defined as $0$ when $\tilde{\bm\psi}^-_j=\bm 0$. Combining the two cases will give~\eqref{eqn:full_angular_measure}.
\end{proof}

\begin{lemma}
\label{lem:mar_survival}
If $\omega\sim \mathcal{S}(\lambda,\kappa,\nu,\delta)$, then as $u \rightarrow 1$,
\begin{align}
\label{eqn:survival}
F_{\omega}^{-1} (u) &= \frac{(1-u)^{-\frac{1}{\lambda}}}{ \nu^{-1} c_{\lambda}^{-\frac{1}{\lambda}} (1+\kappa)^{-\frac{1}{\lambda}}} \left\{ 1+ o(1)\right\}, \\
\label{eqn:cdf}
F_{\omega}^{-1} (1-u) &= -\frac{(1-u)^{-\frac{1}{\lambda}}}{ \nu^{-1} c_{\lambda}^{-\frac{1}{\lambda}} (1-\kappa)^{-\frac{1}{\lambda}}} \left\{ 1+ o(1)\right\}, 
\end{align}
where $c_{\lambda}= \sin(\pi \lambda/2) \Gamma (\lambda)/ \pi$.
\end{lemma}
\begin{proof}
From \citet{nolan2020univariate}, we know that the survival function of a stable random variable with $\lambda \in (0,2)$, $\kappa \in (-1,1)$ satisfies: 
\begin{align*}
\bar{F}_{\omega} (w) = \nu^{\lambda} c_{\lambda} (1+\kappa) w^{-\lambda} (1+ o(1)),\text{ as }w \rightarrow \infty.
\end{align*}
As $u\rightarrow 1$, by definition of quantile function, 
\begin{align*}
1-u &= \bar{F}_{\omega} ( F_{\omega}^{-1}(u))\\
& =\nu^{\lambda} c_{\lambda} (1+ \kappa) \bigg\{ F_{\omega}^{-1}(u)\bigg\}^{-\lambda} \bigg( 1+ o(1) \bigg),
\end{align*}
which gives
\begin{align*}
\bigg\{ F_{\omega}^{-1}(u)\bigg\}^{-\lambda} = \frac{1-u}{ \nu^{\lambda} c_{\lambda} (1+\kappa) (1+o(1))}.
\end{align*}
Raising the power by $-\frac{1}{\lambda}$ on both sides of this equation will give us \eqref{eqn:survival}.

Conversely, we know that the distribution function of a stable random variable satisfies: 
\begin{align*}
F_{\omega} (-w) = \nu^{\lambda} c_{\lambda} (1-\kappa) w^{-\lambda} (1+ o(1)),\text{ as }w \rightarrow \infty.
\end{align*}
As $u\rightarrow 1$, by definition of quantile function, 
\begin{align*}
1-u &= F_{\omega} ( F_{\omega}^{-1}(1-u))\\
& =\nu^{\lambda} c_{\lambda} (1- \kappa) \bigg\{ -F_{\omega}^{-1}(1-u)\bigg\}^{-\lambda} \bigg( 1+ o(1) \bigg),
\end{align*}
which gives
\begin{align*}
\bigg\{ -F_{\omega}^{-1}(1-u)\bigg\}^{-\lambda} = \frac{1-u}{ \nu^{\lambda} c_{\lambda} (1-\kappa) (1+o(1))}.
\end{align*}
Raising the power by $-\frac{1}{\lambda}$ on both sides of this equation will give us \eqref{eqn:cdf}.
\end{proof}

\subsection{The case with the same tail index}\label{sec;same_RV}
Before we prove Theorem~\ref{theorem1}, we introduce the following classic and useful result.
\begin{lemma}[Breiman's lemma, see \citet{Breiman1965,Cline1994} and \citet{Pakes2004}, Lemma~2.1]\label{lem:breiman}
	Suppose $X\sim F$, $Y\sim G$ are independent random variables. If $\overline{F}\in\RV_{-\alpha}^\infty$ with $\alpha\geq 0$ and $Y\geq 0$ with $\E (Y^{\alpha+\varepsilon})<\infty$ for some $\varepsilon>0$, then $$\overline{F}_{XY}(x)\sim \E\left(Y^{\alpha}\right)\, \overline{F}(x), \qquad x\to\infty.$$ 
	Equivalently, if $F\in\mathrm{ET}_{\alpha}$ and $\E(e^{(\alpha+\epsilon)Y})<\infty$, then $\bar{F}_{X+Y}(x)=\overline{F\star G}(x)\sim \E\left(e^{\alpha Y}\right)\, \overline{F}(x)$.
\end{lemma}
\begin{proof}[Proof of Theorem~\ref{theorem1} ]
    Using the stable property, we know that $\psi_{ij}\omega_j\sim\mathcal{S}(\lambda, \sign(\psi_{ij})\kappa_j,|\psi_{ij}|\nu_j, \psi_{ij}\delta_j)$ and thus $X_i\sim\mathcal{S}(\lambda, \bar{\kappa}_i,\bar{\nu}_i, \bar{\delta}_i) $ 
in which
\begin{align*}
    \bar{\kappa}_i &= \frac{\sum_{j=1}^J\sign(\psi_{ij})\kappa_j|\psi_{ij}|^\lambda\nu_j^\lambda}{\sum_{j=1}^J|\psi_{ij}|^\lambda\nu_j^\lambda},\;\bar{\nu}_i=\left(\sum_{j=1}^J|\psi_{ij}|^\lambda\nu_j^\lambda\right)^{1/\lambda},\;
    \bar{\delta}_i=\bar{\delta}_i=\sum_{j=1}^J \psi_{ij}\delta_k,
\end{align*}
for $i=1,\;2$. Therefore, by Lemma~\ref{lem:mar_survival}, 
$$F_{X_i}^{-1} (u) = \frac{(1-u)^{-\frac{1}{\lambda}}}{ \bar{\nu}_i^{-1} c_{\lambda}^{-\frac{1}{\lambda}} (1+\bar{\kappa}_i)^{-\frac{1}{\lambda}}} \left\{ 1+ o(1)\right\}, \textup{\text{ as }} u \rightarrow 1,$$
in which $\bar{\nu}_i^{-1}(1+\bar{\kappa}_i)^{-\frac{1}{\lambda}}=\left\{\sum_{j=1}^J(1+\sign(\psi_{ij})\kappa_j)|\psi_{ij}|^\lambda\nu_j^\lambda\right\}^{-1/\lambda}$.

We first consider the same-tail dependence. We note that
\begin{footnotesize}
    \begin{align*}
    \pr\{F_{X_1}&(X_1)>u, F_{X_2}(X_2)>u\} \\
    &= \pr\left[X_1>\frac{(1-u)^{-\frac{1}{\lambda}}}{ \bar{\nu}_1^{-1} c_{\lambda}^{-\frac{1}{\lambda}} (1+\bar{\kappa}_1)^{-\frac{1}{\lambda}}} \left\{ 1+ o(1)\right\}, X_2>\frac{(1-u)^{-\frac{1}{\lambda}}}{ \bar{\nu}_2^{-1} c_{\lambda}^{-\frac{1}{\lambda}} (1+\bar{\kappa}_2)^{-\frac{1}{\lambda}}} \left\{ 1+ o(1)\right\}\right].
\end{align*}
\end{footnotesize}
To directly apply Expression~\eqref{eqn:full_angular_measure} for this case, we write
\begin{equation}
    \begin{split}
        {\bm\psi}^+_j&=\max\left(\begin{pmatrix}0\\0\end{pmatrix},\begin{pmatrix}
            \frac{\psi_{1j}\nu_j(1+\kappa_j)^{1/\lambda}}{\bar{\nu}_1 c_{\lambda}^{1/\lambda}(1+\bar{\kappa}_1)^{1/\lambda}}\\
            \frac{\psi_{2j}\nu_j(1+\kappa_j)^{1/\lambda}}{\bar{\nu}_2c_{\lambda}^{1/\lambda}(1+\bar{\kappa}_2)^{1/\lambda}}
        \end{pmatrix}\right),\\
        {\bm\psi}^-_j&=-\min\left(\begin{pmatrix}0\\0\end{pmatrix},\begin{pmatrix}
            \frac{\psi_{1j}\nu_j(1-\kappa_j)^{1/\lambda}}{\bar{\nu}_1 c_{\lambda}^{1/\lambda}(1+\bar{\kappa}_1)^{1/\lambda}}\\
            \frac{\psi_{2j}\nu_j(1-\kappa_j)^{1/\lambda}}{\bar{\nu}_2c_{\lambda}^{1/\lambda}(1+\bar{\kappa}_2)^{1/\lambda}}
        \end{pmatrix}\right).
    \end{split}
\end{equation}
Then, under the componentwise min-operator norm, 
\begin{equation*}
    \begin{split}
        \|{\bm\psi}^+_j\| = \begin{cases}
            \min\left( \frac{\psi_{1j}\nu_j(1+\kappa_j)^{1/\lambda}}{\bar{\nu}_1 c_{\lambda}^{1/\lambda}(1+\bar{\kappa}_1)^{1/\lambda}},
            \frac{\psi_{2j}\nu_j(1+\kappa_j)^{1/\lambda}}{\bar{\nu}_2c_{\lambda}^{1/\lambda}(1+\bar{\kappa}_2)^{1/\lambda}}\right),&\text{ if }\psi_{1j}>0, \;\psi_{2j}>0,\\
            0,&\text{ otherwise},
        \end{cases}
    \end{split}
\end{equation*}
and
\begin{equation*}
    \begin{split}
        \|{\bm\psi}^-_j\| = \begin{cases}
            \min\left( \frac{-\psi_{1j}\nu_j(1-\kappa_j)^{1/\lambda}}{\bar{\nu}_1 c_{\lambda}^{1/\lambda}(1+\bar{\kappa}_1)^{1/\lambda}},
            \frac{-\psi_{2j}\nu_j(1-\kappa_j)^{1/\lambda}}{\bar{\nu}_2c_{\lambda}^{1/\lambda}(1+\bar{\kappa}_2)^{1/\lambda}}\right),&\text{ if }\psi_{1j}<0, \;\psi_{2j}<0,\\
            0,&\text{ otherwise}.
        \end{cases}
    \end{split}
\end{equation*}
Then,
\begin{footnotesize}
    \begin{equation*}
    \sum_{j=1}^J \left\|{\bm\psi}^+_j\right\|^\lambda +  \sum_{j=1}^J \left\|{\bm\psi}^-_j\right\|^\lambda=\sum_{\{j:\;\psi_{1j}\psi_{2j}>0\}} \min\left\{\frac{|\psi_{1j}|^\lambda\nu_j^\lambda(1+\sign(\psi_{1j})\kappa_j)}{\bar{\nu}_1^\lambda c_{\lambda}(1+\bar{\kappa}_1)},
            \frac{|\psi_{2j}|^\lambda\nu_j^\lambda(1+\sign(\psi_{2j})\kappa_j)}{\bar{\nu}_2^\lambda c_{\lambda}(1+\bar{\kappa}_2)}\right\}.
\end{equation*}
\end{footnotesize}
Consequently,
\begin{footnotesize}
    \begin{equation*}
       \begin{split}
            \chi^{\rm UU}_{ij} &= \frac{\pr\{F_{X_1}(X_1)>u, F_{X_2}(X_2)>u\}}{1-u}\\
            &=\sum_{\{j:\;\psi_{1j}\psi_{2j}>0\}} \min\left\{ \frac{|\psi_{1j}|^\lambda\nu_j^\lambda(1+\sign(\psi_{1j})\kappa_j)}{\bar{\nu}_1^\lambda (1+\bar{\kappa}_1)},
            \frac{|\psi_{2j}|^\lambda\nu_j^\lambda(1+\sign(\psi_{2j})\kappa_j)}{\bar{\nu}_2^\lambda (1+\bar{\kappa}_2)}\right\}\\
            &=\sum_{\{j:\;\psi_{1j}\psi_{2j}>0\}} \min\left\{\frac{|\psi_{1j}|^\lambda\nu_j^\lambda(1+\sign(\psi_{1j})\kappa_j)}{\sum_{j=1}^J |\psi_{1j}|^\lambda\nu_j^\lambda(1+\sign(\psi_{1j})\kappa_j)},
            \frac{|\psi_{2j}|^\lambda\nu_j^\lambda(1+\sign(\psi_{2j})\kappa_j)}{\sum_{j=1}^J |\psi_{2j}|^\lambda\nu_j^\lambda(1+\sign(\psi_{2j})\kappa_j)}\right\}.
       \end{split}
    \end{equation*}
\end{footnotesize}

For the lower tails, we then look at
\begin{footnotesize}
    \begin{align*}
    \pr\{F_{X_1}(X_1)<1-u, &F_{X_2}(X_2)<1-u\} \\
    &= \pr\left[-X_1>\frac{(1-u)^{-\frac{1}{\lambda}}}{ \bar{\nu}_1^{-1} c_{\lambda}^{-\frac{1}{\lambda}} (1-\bar{\kappa}_1)^{-\frac{1}{\lambda}}} \left\{ 1+ o(1)\right\}, -X_2>\frac{(1-u)^{-\frac{1}{\lambda}}}{ \bar{\nu}_2^{-1} c_{\lambda}^{-\frac{1}{\lambda}} (1-\bar{\kappa}_2)^{-\frac{1}{\lambda}}} \left\{ 1+ o(1)\right\}\right]\\
    &\sim (1-u)\sum_{j=1}^J \min \left\{\frac{(\tilde{\psi}^{-1}_{1j})^\lambda}{\bar{\nu}^\lambda_1(1-\bar{\kappa}_1)}, \frac{(\tilde{\psi}^{-1}_{2j})^\lambda}{\bar{\nu}^\lambda_2(1-\bar{\kappa}_2)}\right\}\\
    &=(1-u)\sum_{j=1}^J \min \left\{\frac{(1-\kappa_j)\psi^\lambda_{1j}\nu^\lambda_j}{\sum_{j=1}^J(1-\kappa_j)\psi_{1j}^\lambda\nu_j^\lambda}, \frac{(1-\kappa_j)\psi^\lambda_{2j}\nu^\lambda_j}{\sum_{j=1}^J(1-\kappa_j)\psi_{2j}^\lambda\nu_j^\lambda}\right\}.
\end{align*}
\end{footnotesize}
Here, $\psi_{ij}\times (-\omega_j)\sim\mathcal{S}(\lambda, -\sign(\psi_{ij})\kappa_j,|\psi_{ij}|\nu_j, \psi_{ij}\delta_j)$ and
\begin{equation}
    \begin{split}
        {\bm\psi}^+_j&=\max\left(\begin{pmatrix}0\\0\end{pmatrix},\begin{pmatrix}
            \frac{\psi_{1j}\nu_j(1+(-\kappa_j))^{1/\lambda}}{\bar{\nu}_1 c_{\lambda}^{1/\lambda}(1-\bar{\kappa}_1)^{1/\lambda}}\\
            \frac{\psi_{2j}\nu_j(1+(-\kappa_j))^{1/\lambda}}{\bar{\nu}_2c_{\lambda}^{1/\lambda}(1-\bar{\kappa}_2)^{1/\lambda}}
        \end{pmatrix}\right),\\
        {\bm\psi}^-_j&=-\min\left(\begin{pmatrix}0\\0\end{pmatrix},\begin{pmatrix}
            \frac{\psi_{1j}\nu_j(1-(-\kappa_j))^{1/\lambda}}{\bar{\nu}_1 c_{\lambda}^{1/\lambda}(1-\bar{\kappa}_1)^{1/\lambda}}\\
            \frac{\psi_{2j}\nu_j(1-(-\kappa_j))^{1/\lambda}}{\bar{\nu}_2c_{\lambda}^{1/\lambda}(1-\bar{\kappa}_2)^{1/\lambda}}
        \end{pmatrix}\right).
    \end{split}
\end{equation}
Then,
\begin{footnotesize}
    \begin{equation*}
    \sum_{j=1}^J \left\|{\bm\psi}^+_j\right\|^\lambda +  \sum_{j=1}^J \left\|{\bm\psi}^-_j\right\|^\lambda=\sum_{\{j:\;\psi_{1j}\psi_{2j}>0\}} \min\left\{\frac{|\psi_{1j}|^\lambda\nu_j^\lambda(1-\sign(\psi_{1j})\kappa_j)}{\bar{\nu}_1^\lambda c_{\lambda}(1-\bar{\kappa}_1)},
            \frac{|\psi_{2j}|^\lambda\nu_j^\lambda(1-\sign(\psi_{2j})\kappa_j)}{\bar{\nu}_2^\lambda c_{\lambda}(1-\bar{\kappa}_2)}\right\}.
\end{equation*}
\end{footnotesize}
Therefore,
\begin{footnotesize}
    \begin{equation*}
       \begin{split}
            \chi^{\rm LL}_{ij} &= \frac{\pr\{F_{X_1}(X_1)<1-u, F_{X_2}(X_2)<1-u\}}{1-u}\\
            &=\sum_{\{j:\;\psi_{1j}\psi_{2j}>0\}} \min\left\{ \frac{|\psi_{1j}|^\lambda\nu_j^\lambda(1-\sign(\psi_{1j})\kappa_j)}{\bar{\nu}_1^\lambda (1-\bar{\kappa}_1)},
            \frac{|\psi_{2j}|^\lambda\nu_j^\lambda(1-\sign(\psi_{2j})\kappa_j)}{\bar{\nu}_2^\lambda (1-\bar{\kappa}_2)}\right\}\\
            &=\sum_{\{j:\;\psi_{1j}\psi_{2j}>0\}} \min\left\{\frac{|\psi_{1j}|^\lambda\nu_j^\lambda(1-\sign(\psi_{1j})\kappa_j)}{\sum_{j=1}^J |\psi_{1j}|^\lambda\nu_j^\lambda(1-\sign(\psi_{1j})\kappa_j)},
            \frac{|\psi_{2j}|^\lambda\nu_j^\lambda(1-\sign(\psi_{2j})\kappa_j)}{\sum_{j=1}^J |\psi_{2j}|^\lambda\nu_j^\lambda(1-\sign(\psi_{2j})\kappa_j)}\right\}.
       \end{split}
    \end{equation*}
\end{footnotesize}

Similarly, for the opposite-tail dependence, we can apply Expression~\eqref{eqn:full_angular_measure} again to get
\begin{footnotesize}
    \begin{equation*}
       \begin{split}
            \chi^{\rm LU}_{ij} &= \frac{\pr\{F_{X_1}(X_1)<1-u, F_{X_2}(X_2)>u\}}{1-u}\\
            &=\sum_{\{j:\;\psi_{1j}\psi_{2j}<0\}} \min\left\{ \frac{|\psi_{1j}|^\lambda\nu_j^\lambda(1-\sign(\psi_{1j})\kappa_j)}{\bar{\nu}_1^\lambda (1-\bar{\kappa}_1)},
            \frac{|\psi_{2j}|^\lambda\nu_j^\lambda(1+\sign(\psi_{2j})\kappa_j)}{\bar{\nu}_2^\lambda (1+\bar{\kappa}_2)}\right\}\\
            &=\sum_{\{j:\;\psi_{1j}\psi_{2j}<0\}} \min\left\{\frac{|\psi_{1j}|^\lambda\nu_j^\lambda(1-\sign(\psi_{1j})\kappa_j)}{\sum_{j=1}^J |\psi_{1j}|^\lambda\nu_j^\lambda(1-\sign(\psi_{1j})\kappa_j)},
            \frac{|\psi_{2j}|^\lambda\nu_j^\lambda(1+\sign(\psi_{2j})\kappa_j)}{\sum_{j=1}^J |\psi_{2j}|^\lambda\nu_j^\lambda(1+\sign(\psi_{2j})\kappa_j)}\right\},
       \end{split}
    \end{equation*}
\end{footnotesize}
and 
\begin{footnotesize}
    \begin{equation*}
       \begin{split}
            \chi^{\rm UL}_{ij} &= \frac{\pr\{F_{X_1}(X_1)>u, F_{X_2}(X_2)<1-u\}}{1-u}\\
            &=\sum_{\{j:\;\psi_{1j}\psi_{2j}<0\}} \min\left\{ \frac{|\psi_{1j}|^\lambda\nu_j^\lambda(1+\sign(\psi_{1j})\kappa_j)}{\bar{\nu}_1^\lambda (1+\bar{\kappa}_1)},
            \frac{|\psi_{2j}|^\lambda\nu_j^\lambda(1-\sign(\psi_{2j})\kappa_j)}{\bar{\nu}_2^\lambda (1-\bar{\kappa}_2)}\right\}\\
            &=\sum_{\{j:\;\psi_{1j}\psi_{2j}<0\}} \min\left\{\frac{|\psi_{1j}|^\lambda\nu_j^\lambda(1+\sign(\psi_{1j})\kappa_j)}{\sum_{j=1}^J |\psi_{1j}|^\lambda\nu_j^\lambda(1+\sign(\psi_{1j})\kappa_j)},
            \frac{|\psi_{2j}|^\lambda\nu_j^\lambda(1-\sign(\psi_{2j})\kappa_j)}{\sum_{j=1}^J |\psi_{2j}|^\lambda\nu_j^\lambda(1-\sign(\psi_{2j})\kappa_j)}\right\}.
       \end{split}
    \end{equation*}
\end{footnotesize}
This concludes the proof for Theorem~\ref{theorem1}.
\end{proof}

\subsection{The case with the different tail indices}\label{sec;diff_RV}
Before we prove Theorem~\ref{theorem2}, we need to introduce another useful result.
\begin{lemma}[Tail properties of convolutions of subexponential (SE) distributions; \citeauthor{Goldie1998}, \citeyear{Goldie1998}]\label{lem:sub-exp}
    If $F\in\mathcal{S}$ and $\overline{G}_j(x)\sim c_j \overline{F}(x)$, $0\leq c_j < \infty$, then $\overline{G_1\star G_2}(x) \sim (c_1+c_2)\overline{F}(x)$.
\end{lemma}

\begin{proof}[Proof of Theorem~\ref{theorem2} ]
    Let $\omega_{1t},\;...,\;\omega_{Jt}$ be a sequence of independent stable variables indexed in time $t$ such that $\omega{jt} \sim \mathcal{S} (\lambda_j,\kappa_j,\nu_j,\delta_j)$, $j=1,...,J$. Without loss of generality, we assume that they are arranged such that $\lambda_1 \leq \lambda_2 \leq \cdots \leq \lambda_J<2$.

Define $\tilde{\lambda} = \min_j \lambda_{j}$, and the potentially multiple indices at location $i$ where the minimum is realized as $\mathcal{T} = \{j: \max_{i=1,2}\psi_{ij}> 0, \lambda_{j} = \lambda\}$. 
If $\mathcal{T}\neq \emptyset$, define
\begin{equation*}
    Y_{i}=\tilde{Y}_{i} + \tilde{Y}^{\rm c}_{i}, \quad \tilde{Y}_{i}=\sum_{j\in \mathcal{T}}\psi_{ij}\omega_{j},\quad \tilde{Y}^{\rm c}_{i}=\sum_{j\in \mathcal{T}^{\rm c}}\psi_{ij}\omega_{j}, \quad i=1,2.
\end{equation*}
Then, the dependence properties of the pair $(\tilde{Y}_{1}, \tilde{Y}_{2})^{\top}$ is covered by the case in Section~\ref{sec;same_RV}. 
The two locations are dominated by the exact same set of most heavy-tailed innovations and other sum terms $\tilde{Y}^{\rm c}_{1}$ and $\tilde{Y}^{\rm c}_{2}$ cannot modify the spectral measure or the $\chi$-coefficient. That is, $\chi(Y_1,Y_2)=\chi(\tilde{Y}_{1}, \tilde{Y}_{2})$. To show this, 
we utilize the fact that regularly varying random variables are a special case of subexponential random variables.

Therefore, we have by Lemma~\ref{lem:sub-exp} that
\begin{equation*}
    \pr(\tilde{Y}_{1}+\tilde{Y}^{\rm c}_{1}>x)\sim (\tilde{c}_{1}+\tilde{c}^{\rm c}_{1})\overline{F}(x)
\end{equation*}
Note the following inequality:
\begin{equation}\label{eqn:sandwich}
    \begin{split}
        \pr(\min(\tilde{Y}_{1},\tilde{Y}_{2})+\min(\tilde{Y}^{\rm c}_{1},\tilde{Y}^{\rm c}_{2})>x)\leq &\pr(\tilde{Y}_{1}+\tilde{Y}^{\rm c}_{1}>x,\tilde{Y}_{2}+\tilde{Y}^{\rm c}_{2}>x)\leq\\ &\pr(\min(\tilde{Y}_{1},\tilde{Y}_{2})+\max(\tilde{Y}^{\rm c}_{1},\tilde{Y}^{\rm c}_{2})>x).
    \end{split}
\end{equation}
Then, by Equation~\eqref{eqn:full_angular_measure},
\begin{align*}
    \pr(\min(\tilde{Y}^{\rm c}_{1},\tilde{Y}^{\rm c}_{2})>x)\sim c_{\min}\overline{F}(x),\text{ where }c_{\min}=0\\
    \pr(\max(\tilde{Y}^{\rm c}_{1},\tilde{Y}^{\rm c}_{2})>x)\sim c_{\min}\overline{F}(x),\text{ where }c_{\max}=0
    \end{align*}
Therefore, 
$\pr(\tilde{Y}_{1}+\tilde{Y}^{\rm c}_{1}>x,\tilde{Y}_{2}+\tilde{Y}^{\rm c}_{2}>x)\sim \pr(\min(\tilde{Y}_{1},\tilde{Y}_{2})>x)$.
\end{proof}

\section{Linear combinations of exponentially-tailed innovations}\label{proof:VG}
If $X\sim\text{VG}(\xi,\theta,\sigma,\mu)$, 
its PDF is specified in Section~\ref{sec:2.1}.
The asymptotic approximations for its tails are presented in \citet{fischer2023}:
\[
\overline{F}(x) \;\sim\;
\frac{1}{2^{\xi/2}\,\bigl(\theta^2 + \sigma^2\bigr)^{\xi/4}\,\lambda^{+}\,\Gamma(\xi/2)}\;
x^{\tfrac{\xi}{2}-1}\,\exp\{-\lambda^{+}\,(x-\mu)\}
\quad \text{as } x \to +\infty,
\]
and
\[
F(x) \;\sim\;
\frac{1}{2^{\xi/2}\,\bigl(\theta^2 + \sigma^2\bigr)^{\xi/4}\,\lambda^{-}\,\Gamma(\xi/2)}\;
\bigl(-x\bigr)^{\tfrac{\xi}{2}-1}\,\exp\{\lambda^{-}\,(x-\mu)\}
\quad \text{as } x \to -\infty,
\]
in which $\lambda^{+}
\;:=\;
\frac{\sqrt{\theta^2 + \sigma^2} - \theta}{\sigma^2}$ and $\lambda^{-}
\;:=\;
\frac{\sqrt{\theta^2 + \sigma^2} + \theta}{\sigma^2}$. Thus, $\overline{F}\in \text{ET}_{\lambda^{+}}$ and $F(-\cdot)\in \text{ET}_{\lambda^{-}}$.


\begin{proof}[Proof of Theorem~\ref{thm:ET} ]
Consider again
\begin{equation*}
    \begin{cases} X_{1}&=\psi_{11}\omega_{1}+\ldots +\psi_{1J}\omega_{J},\\
X_{2}&=\psi_{21}\omega_{1}+\ldots +\psi_{2J}\omega_{J},
    \end{cases}
\end{equation*}
and assume that the independent variables $\omega_j$ are in $\mathrm{ET}^\infty_{\lambda_j^+}$ (upper exponential tail with rate parameter $\lambda_j^+$) and $\mathrm{ET}^{-\infty}_{\lambda_j^-}$ (lower exponential tail with rate parameter $\lambda_j^-$), where we use the convention that a variable with tail lighter than exponential has $\lambda^\pm_j=\infty$.
Set 
\begin{equation*}
    \bm\psi_{j}^{UU}=\max\left\{\begin{pmatrix}
            \psi_{1j}\\\psi_{2j}
        \end{pmatrix}, \begin{pmatrix}
            0\\0
        \end{pmatrix}\right\}/\lambda^+_j+\max\left\{\begin{pmatrix}
            -\psi_{1j}\\-\psi_{2j}
        \end{pmatrix}, \begin{pmatrix}
            0\\0
        \end{pmatrix}\right\}/\lambda^-_j,
\end{equation*}
whose elements capture the scale parameter (i.e., inverse rate parameter) of the upper tail of $\psi_{1j}\omega_{j}$ and $\psi_{2j}\omega_{j}$ when $\psi_{1j}\psi_{2j}>0$. Define the maximum scales as $\tilde{\psi}_{i}^{UU} = \max_j \psi^{UU}_{ij}$, $i=1,2$, and the potentially multiple indices where the maximum is realized as $I^{UU}_i = \{j: \psi_{ij} = \tilde{\psi}_{i}^{UU}\}$.

Similarly, we set
\begin{equation*}
    \begin{split}
        \bm\psi_{j}^{UL}&=\max\left\{\begin{pmatrix}
            \psi_{1j}\\-\psi_{2j}
        \end{pmatrix}, \begin{pmatrix}
            0\\0
        \end{pmatrix}\right\}/\lambda^+_j+\max\left\{\begin{pmatrix}
            -\psi_{1j}\\\psi_{2j}
        \end{pmatrix}, \begin{pmatrix}
            0\\0
        \end{pmatrix}\right\}/\lambda^-_j,\\
        \bm\psi_{j}^{LU}&=\max\left\{\begin{pmatrix}
            -\psi_{1j}\\\psi_{2j}
        \end{pmatrix}, \begin{pmatrix}
            0\\0
        \end{pmatrix}\right\}/\lambda^+_j+\max\left\{\begin{pmatrix}
            \psi_{1j}\\-\psi_{2j}
        \end{pmatrix}, \begin{pmatrix}
            0\\0
        \end{pmatrix}\right\}/\lambda^-_j.\\
         \bm\psi_{j}^{LL}&=\max\left\{\begin{pmatrix}
            -\psi_{1j}\\-\psi_{2j}
        \end{pmatrix}, \begin{pmatrix}
            0\\0
        \end{pmatrix}\right\}/\lambda^+_j+\max\left\{\begin{pmatrix}
            \psi_{1j}\\\psi_{2j}
        \end{pmatrix}, \begin{pmatrix}
            0\\0
        \end{pmatrix}\right\}/\lambda^-_j.
    \end{split}
\end{equation*}
Also define the maximum scales as $\tilde{\psi}_{i}^{LL}$, $\tilde{\psi}_{i}^{UL}$ and $\tilde{\psi}_{i}^{LU}$, and the maximum index sets $I^{LL}_i$, $I^{UL}_i$ and $I^{LU}_i$, $i=1,2$.

We only show the cases when $A=UU$ and $A=UL$. The proofs for the other two cases are highly analogous. 

When $A=UU$, if $I^{UU}_1=I^{UU}_2$ and $\tilde{\psi}_{i}^{UU}>0$ for $i=1,2$, then the two variables $\sum_{j \in I^{UU}_i} \omega_{j} \psi_{ij}/\tilde{\psi}_{i}^{UU}$ are identical for $i=1,2$ and exponential-tailed at $\infty$ with scale $1$. Moreover, using Breiman's lemma with the previous variable representing the heavier-tailed component, we find that the marginal distributions of $X_i/\tilde{\psi}_{i}^{UU}-\log m^{UU}_i$ are tail-equivalent with $\Pr(X_i/\tilde{\psi}_{i}^{UU}-\log m^{UU}_i>x) \sim m^{UU}_i r(x+\log m^{UU}_i)\exp(-(x+\log m^{UU}_i))\sim r(x)\exp(-x)$, in which $r\in \text{RV}_\lambda^\infty$. Therefore, we can calculate $\chi_{12}^{UU}$ as 
    \begin{align*}
        \chi_{12}^{UU} & = \lim_{x\rightarrow\infty}\frac{\Pr(X_1/\tilde{\psi}_1^{UU}-\log m^{UU}_1>x, X_2/\tilde{\psi}_2^{UU}-\log m^{UU}_2>x)}{\Pr(X_1/\tilde{\psi}_1^{UU}-\log m^{UU}_1>x)}\\
        &=\lim_{x\rightarrow\infty} \frac{\Pr\left(\sum_{j \in I^{UU}_i} \omega_{j} \psi_{ij}/\tilde{\psi}_{i}^{UU} + \min_{i=1,2}\left(\sum_{j \not\in I^{UU}_i} \omega_{j} \psi_{ij}/\tilde{\psi}_{i}^{UU}-\log m^{UU}_i \right)>x\right)}{\Pr(X_1/\tilde{\psi}_1^{UU}-\log m^{UU}_1>x)},
    \end{align*}
and applying Breiman's lemma both to the numerator and denominator gives the formula for $\chi_{12}^{UU}$.

When $A=UL$, $I^{UL}_1=I^{UL}_2$ and $\tilde{\psi}_{i}^{UL}>0$ for $i=1,2$, then the two variables $\sum_{j\in I^{UL}_2} \omega_{j}(\psi_{1j}/\tilde{\psi}_{1}^{UL}, -\psi_{2j}/\tilde{\psi}_{2}^{UL})^{\top}$ are identical for $i=1,2$ and exponential-tailed at $\infty$ with scale $1$. Similarly,  Breiman's lemma gives us
\begin{equation*}
    \Pr(X_1/\tilde{\psi}_{1}^{UL}-\log m^{UL}_1>x) \sim m^{UL}_1 r(x+\log m^{UL}_1)\exp(-(x+\log m^{UL}_1))\sim r(x)\exp(-x)
\end{equation*}
and
\begin{equation*}
    \begin{split}
        \Pr(X_2/\tilde{\psi}_{2}^{UL}&+\log m^{UL}_2<-x) =
        \Pr(-X_2/\tilde{\psi}_{2}^{UL}-\log m^{UL}_2 > x) \\
        &\sim m^{UL}_2 r(x+\log m^{UL}_2)\exp(-(x+\log m^{UL}_2))\sim r(x)\exp(-x).       
    \end{split}
\end{equation*}

Therefore, we can calculate $\chi_{12}^{UL}$ as 
    \begin{align*}
        \chi_{12}^{UL} & = \lim_{x\rightarrow\infty}\frac{\Pr(X_1/\tilde{\psi}_1^{UL}-\log m^{UL}_1>x, X_2/\tilde{\psi}_2^{UL}+\log m^{UL}_2< - x)}{\Pr(X_1/\tilde{\psi}_1^{UL}-\log m^{UL}_1>x)}\\
        &=\lim_{x\rightarrow\infty} \frac{\Pr\left(\sum_{j \in I^{UL}_1} \omega_{j} \psi_{1j}/\tilde{\psi}_{1}^{UL} + \min_{i=1,2}\left(\tilde{Y}^{UL}_i-\log m^{UL}_i \right)>x\right)}{\Pr(X_1/\tilde{\psi}_1^{UL}-\log m^{UL}_1>x)},
    \end{align*}
and applying Breiman's lemma both to the numerator and denominator gives the formula for $\chi_{12}^{UL}$.
\end{proof}

\section{Long-term dependence}
\begin{proof}[Proof of Theorem~\ref{thm:long_term} ]
    Since the transition matrix $\bm{M}$ is non-defective, we can take the eigen-decomposition on the transition matrix to get $\bm{M}=\bm{Q} \boldsymbol{\Lambda} \bm{Q}^{-1}$, in which $\bm{Q}\in \mathbb{R}^{K\times K}$ whose $i^{\rm th}$ column is the eigenvector $\bm q_i$ of $\bm{M}$ and $\boldsymbol{\Lambda}=\mathrm{diag}\{\lambda_1, \ldots, \lambda_{K}\}$ contains all eigenvalues of $\bm{M}$.
Then, we have $\bm{M}^t = \bm{Q} \boldsymbol{\Lambda}^t \bm{Q}^{-1},\; t \in \mathbb{N}^+$.

Given the spectral radius $\|\bm{M}\|=\max\{|\lambda_i|:i=1,\ldots, K\}<1$, all eigenvalues satisfy $|\lambda_i|<1$ and $\lambda_i^t\rightarrow 0$ as $t\rightarrow\infty$.
Thus, $\bm{M}^t = \bm{Q} \boldsymbol{\Lambda}^t \bm{Q}^{-1}\rightarrow \bm{0}$, $t\rightarrow\infty$. 

Therefore, for a fixed time $t_0$, we can examine the long-term effect of $\bm Y_{t_0}$ on $\bm Y_{t_0+\Delta t}$ as $\Delta t\rightarrow \infty$. Note that
\begin{align*}
    \bm Y_{t_0} &= \sum_{r=0}^{t_0} \bm\Psi \bm{M}^{{t_0}-r}\bm\omega_r+\bm\varepsilon_t,\\
    \bm Y_{{t_0}+\Delta t} &=\sum_{r=0}^{t_0} \bm\Psi \bm{M}^{{t_0}+\Delta t-r} \bm\omega_r + \sum_{r={t_0}+1}^{{t_0}+\Delta t} \bm\Psi \bm{M}^{{t_0}+\Delta t-r} \bm\omega_r+\bm\varepsilon_{{t_0}+\Delta t}.
\end{align*}

Consequently, the first term on the right‑hand side of the previous display converges to zero almost surely as $\Delta t\rightarrow\infty$, and thus $\bm Y_{t_0}$ and $\bm Y_{t_0+\Delta t}$ are asymptotically independent.

To see how the state at time $t_0$ influences the state at $t+\Delta t$, we write the innovations in the eigenbasis of $\bm M$. Let
$$ \boldsymbol{\omega}_r = \bm{Q} \bm{c}_r = \sum_{k=1}^{K} c_{kr} \bm Q_{\cdot,k}$$ for some random vectors
$\bm{c}_r = (c_{1r}, \ldots, c_{Kr})^{\top}$, $r=0, \ldots, t_0$. Then, it follows that 
\[
\bm{M}^{{t_0}+\Delta t-r} \bm\omega_r = \bm{Q} \boldsymbol{\Lambda}^{t_0+\Delta t} \bm{c}_r 
= \sum_{k=1}^{K} c_{kr} \lambda_k^{t_0+\Delta t-r} \bm Q_{\cdot,k}, 
\qquad r = 0, 1, \ldots, t_0. 
\]

Therefore, the influence of 
 $\bm Y_{t_0}$ on $\bm Y_{t_0+\Delta t}$ hinges on the magnitudes of the eigenvalues $\{\lambda_k\}$: if $|\lambda_k| < 1$, then $\lim_{t \to \infty} \lambda_k^t = 0$, and thus the $k^{\rm th}$ term in the linear expansion of $\bm{M}^{{t_0}+\Delta t-r}\bm \omega_r$ goes to zero. More compactly, when the spectral radius satisfies $\|\bm{M}\| < 1$, it follows that 
$$\lim_{\Delta t \to \infty} \sum_{r=0}^{t_0} \bm\Psi \bm{M}^{{t_0}+\Delta t-r} \bm\omega_r  = \bm{0} \text{  almost surely}$$  
for $r = 0, 1, \ldots, t_0$. Thus, the zero vector acts as a stable equilibrium.  In summary:
\begin{enumerate}[(i)]
    \item \quad If $\lambda_k$ is real and $\lambda_k > 1$, then the $k^{\rm th}$ component of $\sum_{r=0}^{t_0} \bm\Psi \bm{M}^{{t_0}+\Delta t-r} \bm\omega_r$ in the expansion exhibits exponential growth.
  \item \quad If $\lambda_k$ is real and $0 < \lambda_k < 1$, then the $k^{\rm th}$ component of $\sum_{r=0}^{t_0} \bm\Psi \bm{M}^{{t_0}+\Delta t-r} \bm\omega_r$ in the expansion exhibits exponential decay.
\end{enumerate}
Hence stability, and the resulting long-term AI, requires every eigenvalue of $\bm M$ to lie strictly inside the unit circle.
\end{proof}

\section{Prior model details}\label{sec:prior_model}
 When using either the variance-gamma or stable distribution in a regime-switching process, the prior distributions for parameters common to both are specified as:
\begin{align}
\sigma^2_d &\sim \text{Inv.gamma}(r_d,\gamma_d), & 
b(s^*_{k,1},s^*_{k,2}) &\sim \text{Gau}(\bm{0},C_{b}), & 
c &\sim \text{Gau}(\bm{0},C_{c}) \nonumber \\
d &\sim \text{Gau}(\bm{0},C_{d}), & 
e &\sim \text{Gau}(\bm{0},C_{e}), & 
f &\sim \text{Gau}(\bm{0},C_{f}) \label{eq:center_number}\\
\bm{\beta}_k &\sim \text{Gau}(\bm{0},C_{\beta} \bm{I}), & 
\bm{\alpha}_0 &\sim \text{Gau}(\bm{0},C_{\alpha} \bm{I}), & 
\sigma^2_{\text{Gau},k} &\sim \text{Inv.gamma}(r_{g},\gamma_{g}). \nonumber
\end{align}
Specific prior distributions for the variance-gamma and stable distributions are:
\begin{align}
\sigma_{\text{vg},k} &\sim \text{Inv.gamma}(r_{\sigma},\gamma_{\sigma}), &
\theta_{\text{vg},k} &\sim \text{Gau}(\mu_{\theta},C_{\theta}), &
\lambda_{k} &\sim \text{TN}(\mu_{\lambda},\sigma_{\lambda},\ell=0,u=2), \nonumber \nonumber \\
\kappa_{j} &\sim \text{Unif}(-1,1), &
\nu_{s,k} &\sim \text{Inv.gamma}(r_{\nu},\gamma_{\nu}), &
\label{priors}
\end{align}
where $\text{TN}$ denotes the truncated normal distribution with mean $\mu_{\lambda}$ and standard deviation $\sigma_{\lambda}$, bounded by lower and upper limits $\ell$ and $u$. The priors in \eqref{priors} can be set to be weakly informative, letting the heavy-tailed distribution capture extremes while the light-tailed distribution explains the bulk. For instance, centering $\lambda_{k}$ at $\mu_{\lambda} < 2$ encourages heavier tails than a Gaussian. Similarly, setting the mode or mean of $\sigma_{\text{vg},k}$ and $\nu_{s,k}$ higher than that of $\sigma^2_{\text{Gau},k}$ encourages the light-tailed distribution to better capture non-extreme events.

\section{Parameter estimation for the stable distribution with a nuisance term}\label{sec:stable_estimate}
\begin{figure}[t]
\centering
\includegraphics[width=0.8\linewidth]{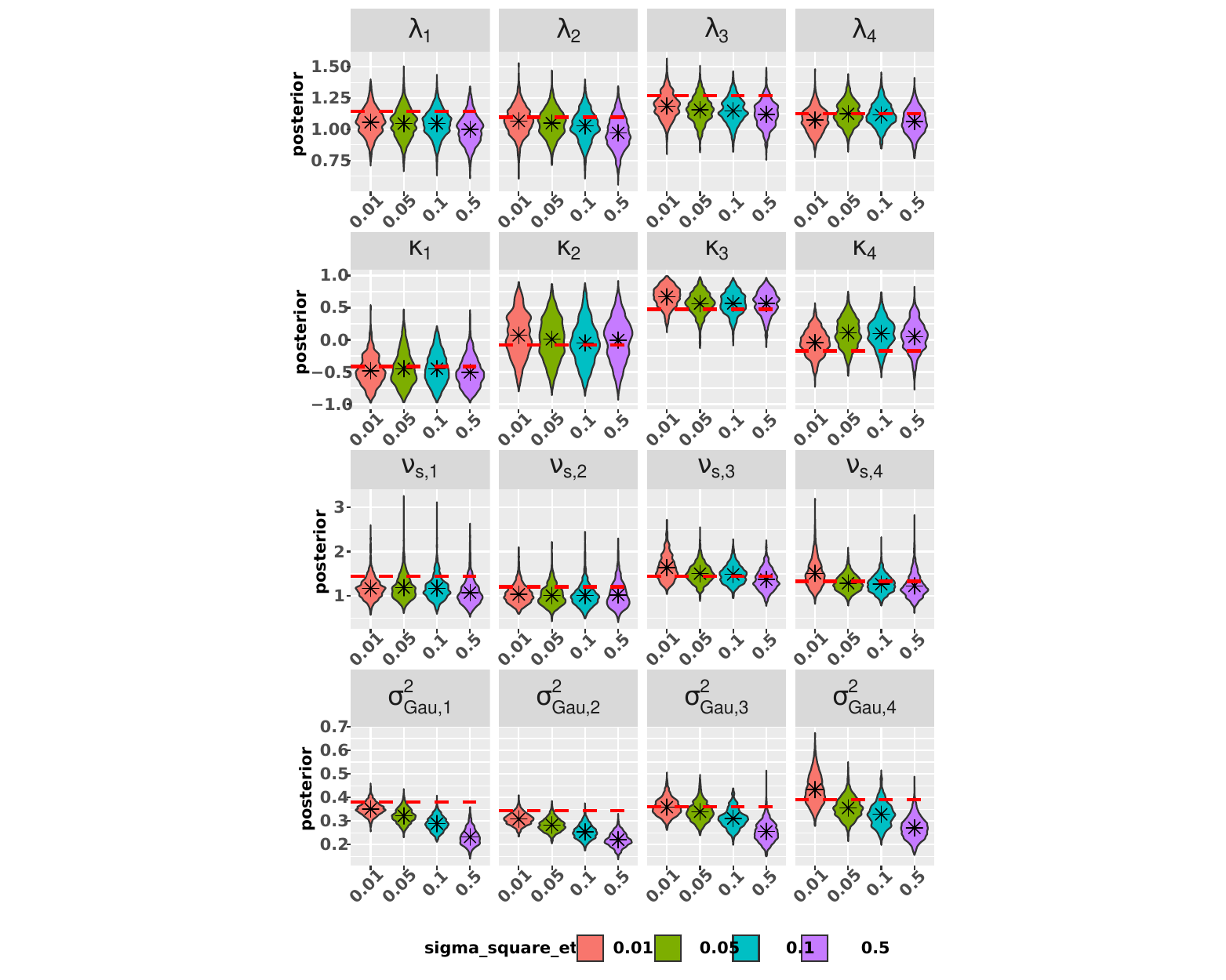}
  \caption{Violin plots of $\lambda_{S,k}$, $\kappa_{S,k}$, $\nu_{S,k}$, and $\sigma^2_{\text{Gau},k}$ for $k=1,...,4$ (from left to right), correspndong to different values of $\sigma^2_{\eta}$. The red dashed horizontal lines represent the true values.}
    \label{stable_estimate}
\end{figure}
We also examine parameter estimation for the stable distribution case with the nuisance term. we generate the innovation $\bm{\omega}_t$ as:
\begin{align}
\label{eqn:4.3.14}
\pi(\omega_{k,t}) &=\begin{cases}
\pi(\text{Gau}(0,\sigma^2_{\text{Gau},k})) & if \quad k\in (1,2), t\leq 110 \\
\pi(\bm{S}(\lambda_{S,k},\kappa_{S,k},\nu_{S,k}=0))  & if \quad k\in (1,2), t> 110 \\
\pi(\text{Gau}(0,\sigma^2_{\text{Gau},k})) & if \quad k\in (3,4), t\leq 60 \\
\pi(\bm{S}(\lambda_{S,k},\kappa_{S,k},\nu_{S,k},0))  & if \quad k\in (3,4), t> 60.
\end{cases}
\end{align}
Accordingly, we generate the basis function coefficients $\bm{\alpha}_t$ using the same number of knots and the matrix $\bm{M}$, as detailed in Section \ref{sec:5}.

Figure \ref{stable_estimate} shows that $\sigma^2_{\text{Gau},k}$ is more sensitive to $\sigma^2_{\eta}$, whereas the other parameters in the stable distribution are more robust. It can be seen that the true parameters are recovered when $\sigma^2_{\eta}$ is small.

\begin{figure}[t]
\centering
\includegraphics[width=1\linewidth]{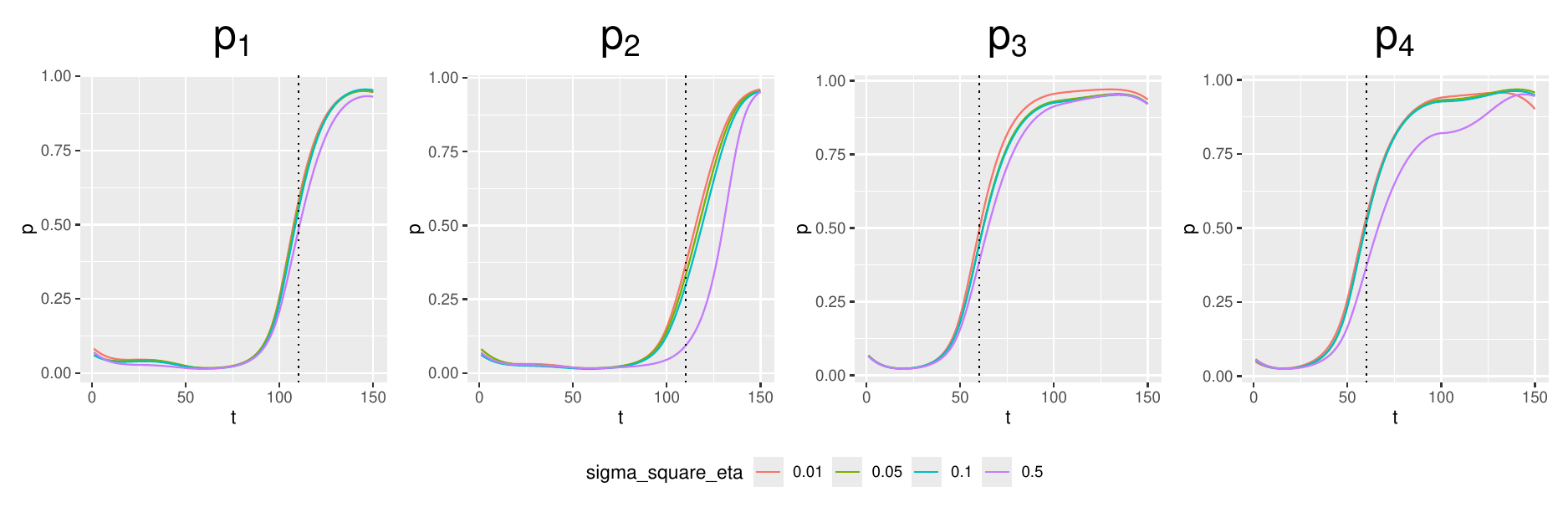}
  \caption{Posterior mean of $p_{kt}$ for $k=1,...,4$ (from left to right) and $t=1,...,T$ with varying $\sigma^2_{\eta}$}
    \label{simul_prob_stable}
\end{figure}

We also check the effect of $\bm{\eta}_t$ on estimating $\bm{p}_t$ (Figure \ref{simul_prob_stable}). As in the variance gamma case, the posterior mean of $\bm{p}_t$ captures the data generating process.

\bibliographystyle{apalike}  
\bibliography{arXiv/references}

\end{document}